\documentclass[twocolumn]{article}

\usepackage{content/preamble}

\usepackage[english]{babel}
\usepackage[sorting=none, style=nature, doi=true, backend=bibtex]{biblatex}

\title{\TITLE}

\begin{document}

\newgeometry{onecolumn, margin=20mm}

\maketitle

\begin{abstract}
  Recent years have seen a rapid rise of \glsentrylongpl{ANN} being employed in a number of cognitive tasks.
The ever-increasing computing requirements of these structures have contributed to a desire for novel technologies and paradigms, including memristor-based hardware accelerators.
Solutions based on memristive crossbars and analog data processing promise to improve the overall energy efficiency.
However, memristor nonidealities can lead to the degradation of neural network accuracy, while the attempts to mitigate these negative effects often introduce design trade-offs, such as those between power and reliability.
In this work, we design nonideality-aware training of memristor-based neural networks capable of dealing with the most common device nonidealities.
We demonstrate the feasibility of using high-resistance devices that exhibit high \IV\ nonlinearity---by analyzing experimental data and employing nonideality-aware training, we estimate that the energy efficiency of memristive vector-matrix multipliers is improved by almost three orders of magnitude (\SI{0.715}{\tera OP\second^{-1}\watt^{-1}} to \SI{381}{\tera OP\second^{-1}\watt^{-1}}) while maintaining similar accuracy.
We show that associating the parameters of neural networks with individual memristors allows to bias these devices toward less conductive states through regularization of the corresponding optimization problem, while modifying the validation procedure leads to more reliable estimates of performance.
We demonstrate the universality and robustness of our approach when dealing with a wide range of nonidealities.

\end{abstract}

\thispagestyle{empty}

\restoregeometry

\section{Introduction}

\Glspl{ANN} are now routinely used in machine learning tasks ranging from text generation~\cite{BrMa2020} to autonomous driving~\cite{FaHo2018}.
However, rapidly increasing number of parameters in modern \glspl{ANN} is making them time- and power-consuming during both training~\cite{StGa2019} and inference~\cite{HaMa2016} stages.
This makes it challenging to apply machine learning approaches in environments where resources are tightly constrained~\cite{WuIa2017, MeKe2021}.

One of the proposed solutions to improve the efficiency of \glspl{ANN} has been to adopt different computer architectures.
The need to transfer data between memory and computing units in the von Neumann architecture is the main bottleneck in modern computers~\cite{Mc2004}; this is especially evident in machine learning where large amounts of data are utilized.
In this specific case, an alternative can be memristor-based \glspl{ANN}, or \glspl{MNN}.
With this approach, memristive crossbar arrays are used to physically compute vector-matrix products, which are one of the most fundamental operations in \glspl{ANN}~\cite{YaSt2013, IeWa2021}.
This is done without the need to constantly move large amounts of data: matrix entries are encoded into memristor conductances, vector entries---into applied voltages, and the result of the operation is extracted from the output currents produced according to Ohm's law and Kirchhoff's current law~\cite{YaSt2013}. 

Memristors---when used and programmed as analog devices---can encode values at much higher density, but at a cost of lower precision.
A number of nonidealities may prevent from accurately programming device conductances or may cause deviations from intended electrical behavior.
Such nonidealities include stuck devices, \gls{D2D} and cycle-to-cycle variability, drift in resistance states, line resistance, \IV\ nonlinearity, and others~\cite{MeJo2019}.

Potential solutions do exist but many of them introduce a number of trade-offs.
For example, to ensure more linear \IV\ characteristics, one may use low-resistance devices~\cite{YaZh2012}; however, this results in higher power consumption.
Alternatively, the effects of \IV\ nonlinearities may be minimized by utilizing pulse-width modulation~\cite{AmAl2020}, but that comes at a cost of increased clock cycles for each encoded input~\cite{CaCo2019}.

Other techniques of dealing with memristor nonidealities include
\begin{itemize}
  \item \textit{in-situ} (re)training of weights (or just a subset of them~\cite{YaWu2020}) to recover from the effects of nonidealities~\cite{ChLi2017, JaRa2019, WaWu2020, LiHu2017, LiYa2019}, including in \glspl{CNN}~\cite{YaWu2020, WaLi2019a}, recurrent structures~\cite{WaLi2019a}, and in neural networks used for reinforcement learning~\cite{WaLi2019}
\item modifying device structure, including inserting a buffer layer~\cite{FaYu2018}, inserting an electro-thermal modulation layer~\cite{WuWu2018}, and adopting bilayer structure~\cite{WoMo2016}
  \item using additional circuitry~\cite{LiHu2018, AmNa2018} to ensure more stable behavior
\end{itemize}
However, many of these approaches are technology-specific and thus difficult to apply to different types of devices.

When optimizing the performance of memristive systems (as opposed to individual devices), software approaches may be preferable because they are usually technology-agnostic.
For general applications, mapping or redundancy schemes can be used to mitigate the effects of faulty-devices~\cite{XiHu2017} or line resistance~\cite{HuSt2016}.
In the specific context of \glspl{MNN}, multiple smaller nonideal networks may replace a large one and increase the accuracy in this way~\cite{JoFr2020}.
Alternatively, modifying \textit{ex-situ} training has been proposed: altering the cost function~\cite{ZhZh2020} or injecting noise into the synaptic weights~\cite{JoLe2020} can make \glspl{MNN} more robust to the effects of nonidealities.

\begin{figure*}[h!]
  \begin{center}
    \includegraphics{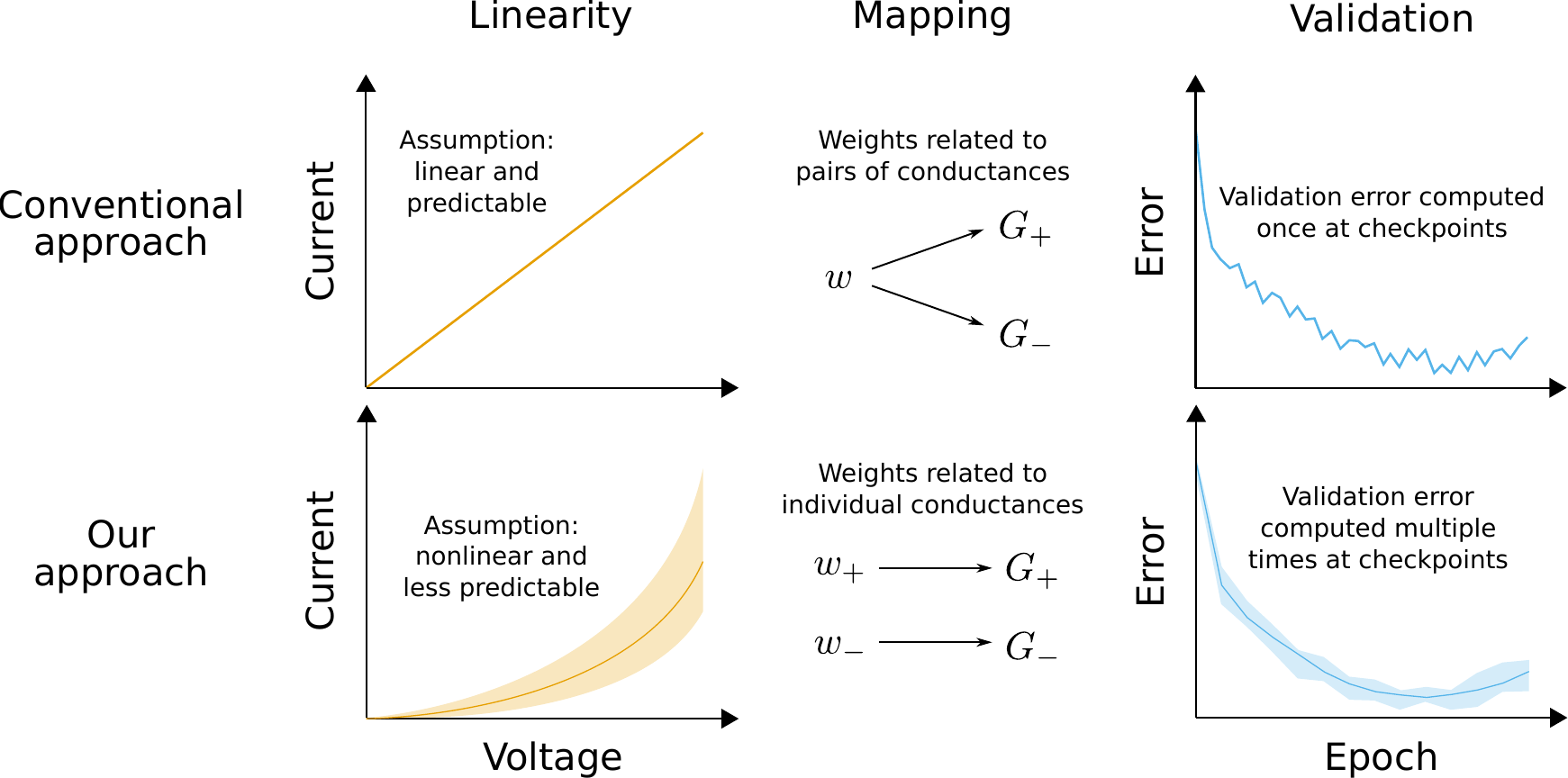}
  \end{center}
  \mycaption{Overview of the novel \textit{ex-situ} training technique designed for \glsentrylongpl{MNN}}{%
    Redefining the output operation of synaptic layers allows to take non-ohmic device behavior into account, which makes working with power-efficient high-resistance (and high-nonlinearity) devices feasible.
    Relating weight parameters in \glsentrylongpl{ANN} to individual conductances in crossbar arrays enables to further decrease power consumption through regularization, as well as to adapt to nonidealities that depend on the conductance values.
    Using an aggregate validation metric provides a more reliable way of assessing \glsentrylong{MNN} performance.
  }\label{fig:overview}
\end{figure*}

Memristor-oriented \textit{ex-situ} training is indeed a very promising method of making \glspl{MNN} feasible.
However, it has been applied by considering only a limited number of nonidealities, while the robustness of this technique is not well understood.
In this work, we propose a number of improvements to memristive \textit{ex-situ} training, which are summarized in \cref{fig:overview}.

Firstly, our novel training technique addresses the problem of nonlinearity.
Deviations from the linear relationship between inputs and outputs in crossbar arrays are a major obstacle, however none of the aforementioned methods directly address this issue.
Although existing works often take \emph{conductance} deviations into account during \textit{ex-situ} training, crossbar arrays are still usually modeled as structures computing a product of a vector of voltages and a matrix of conductances.
To reflect non-ohmic behavior of memristive devices (illustrated in \cref{fig:overview}) during training, we propose incorporating nonlinearities into the node functions of \glspl{MNN}.
Our aim is to embrace memristor nonlinearity so that the network can learn to be robust toward this nonideality---or even take advantage of it---during training based on stochastic gradient descent.
Existing works attempting to take device variations into account during \textit{ex-situ} training often result in potentially lower training accuracy~\cite{LiLi2015a}.
Depending on the nature of the nonideality that \glspl{MNN} are trained on, our method could potentially perform even \emph{better} than conventional \glspl{ANN}.
For example, if nonlinear response of the device is sufficiently consistent, nonlinearities may increase the network capacity~\cite{ZaPe2021}.

We also show how to improve nonideality-aware training by exposing conductances to the training process in a more direct way.
By constraining \gls{MNN} weights to be nonnegative, they can be related to conductances in a linear way.
This allows to
\begin{itemize}
  \item minimize the effects of nonidealities in cases where their severity is dependent on the conductance values of the memristors
  \item employ regularization as a high-level tool for controlling power consumption of \glspl{MNN}
\end{itemize}

Finally, we propose improving validation techniques applied during the training of \glspl{MNN}.
Most memristive nonidealities are stochastic in nature, therefore computing validation metrics only once at specified checkpoints may provide unreliable estimates of neural network performance, as illustrated in \cref{fig:overview}.
We propose computing validation metrics multiple times and using an aggregate value (like the median) to determine which version of the \gls{MNN} to save for inference.

In this work, we explore how to make nonideality-aware training more effective.
We employ experimental data from a \ce{SiO_x} memristor device and a $128 \times 64$ \ce{Ta}/\ce{HfO2} memristor crossbar array, and consider multiple nonidealities: \IV\ nonlinearity, stuck devices, and \gls{D2D} variability.
We give special focus to \IV\ nonlinearity because existing works often prefer low-resistance devices as they exhibit more linear \IV\ behavior.
We show that our proposed training function enables the use of the more power-efficient high-resistance memristors by minimizing the accuracy loss due to high nonlinearity and variability, which typically manifest themselves in high-resistance devices~\cite{MeMu2017}.
We demonstrate how the proposed mapping between nonnegative weights and memristor conductances enables further power savings through $\ell _1$ regularization.
We additionally show how \gls{MNN} validation during training may be improved by taking the stochastic nature of many of the nonidealities into account.

Importantly, we demonstrate the feasibility of our novel training technique.
Networks produced using nonideality-aware \text{ex-situ} training are better adapted to nonidealities and can even handle significant uncertainty in device behavior or designers' understanding of what that behavior is.
Weights learned \emph{once} on a digital system can be transferred onto \emph{multiple} crossbar-based physical implementations, even if the nonidealities manifest themselves differently in each system---our methodology does not require retraining once the weights are mapped onto crossbars.
To assess the robustness of nonideality-aware training, we even expose \glspl{MNN} to different setups than they were trained for---we find that the networks employing this technique are much more robust than the networks trained using the previous, more conventional approaches.

\section{Design}

\subsection{Nonlinearity-Aware Training}

The main focus of this work is the design of a novel \textit{ex-situ} \gls{MNN} training scheme that could handle nonidealities characterized by the nonlinear relationship between inputs and outputs.
Nonidealities like stuck devices, programming inaccuracies or random telegraph noise can typically be represented with a change in memristor conductance \emph{alone}.
They are thus more straightforward to take into consideration during \textit{ex-situ} training because one can inject noise into the conductance array to represent their effect.
We refer to these nonidealities as \textbf{linearity preserving}.
On the other hand, nonidealities like \IV\ nonlinearity or line resistance cannot be represented by simply disturbing the conductances of crossbar devices.
We refer to such nonidealities as \textbf{linearity nonpreserving}.
In this work, we redefine the output operations of the synaptic layers to take the effect of linearity-nonpreserving nonidealities into account.
To the best of our knowledge, this is the first time these nonidealities are addressed during \textit{ex-situ} training. 

Existing works usually model memristor crossbars as structures computing linear dot products, while only activation functions are assumed to introduce nonlinearities.
Specifically, outputs $y_j \in \matr{y} \in \mathbb{R}^{1 \times N}$ are calculated using inputs $x_i \in \matr{x} \in \mathbb{R}^{1 \times M}$, weights $w_{ij} \in \matr{w} \in \mathbb{R}^{M \times N}$, and a nonlinear activation function $f$, as shown in \cref{eq:synapse-normal}.
During inference, $\matr{x}$, $\matr{w}$ and $\matr{y}$ are then mapped onto and from voltages, conductances and currents, respectively.
However, this creates a discrepancy between the linear dot-product training nodes and memristor nonidealities that deviate from ohmic device behavior.

\begin{equation}\label{eq:synapse-normal}
  y_j = f{\left(\sum_{i = 1}^M {x_i w_{ij}} \right)}
\end{equation}

Therefore, we suggest modifying the output operation of the synaptic layers to reflect the nature of linearity-nonpreserving nonidealities.
Specifically, in cases where the nonlinearity is limited to individual devices (i.e.\ where devices experience \IV\ nonlinearity), we propose replacing the approach in \cref{eq:synapse-normal} with the approach in \cref{eq:synapse-nonlinearity-aware}.
That is, the activation function is unchanged but the product is replaced with a nonlinear function $g$ that captures memristors' non-ohmic \IV\ behavior.

\begin{equation}\label{eq:synapse-nonlinearity-aware}
  y_j = f{\left(\sum_{i = 1}^M {g \left(x_i, w_{ij} \right)} \right)}
\end{equation}

The exact form of $g$ will depend on
\begin{itemize}
  \item the mapping scheme, examples of which are explored in \cref{subsec:double-weights-and-mapping}
  \item non-ohmic behavior model, which might typically be determined by the devices used; in \cref{subsec:nonidealities}, we present one possibility motivated by experimental device data
\end{itemize}

\subsection{Modified Weight Implementation}\label{subsec:double-weights-and-mapping}

As mentioned previously, synaptic weights of \glspl{MNN} are typically implemented using conductances of memristive devices.
While weights can usually take any real value, conductances are nonnegative---this presents design challenges.
In this work, we propose training double the number of weights\footnote{The number of memristive devices remains the same as with conventional weight implementation schemes.}, but constrain them to be nonnegative and associate them with individual devices.
This approach creates a more natural mapping between neural network parameters and conductances, as well as enables weight regularization to act as a method for decreasing the power consumption.

\subsubsection{Conventional approaches}

In typical \gls{MNN} implementations, both inputs $x \in \matr{x}$ and outputs $y \in \matr{y}$ are mapped onto voltages\footnote{If the inputs are digital, they may be applied in the form of voltage using digital-to-analog converters or, for example, by utilising binarized inputs~\cite{XuCh2021}, which eliminate the need for such converters altogether.} $V$ and from currents $I$, respectively, in a proportional way using scaling factors $k_V$ and $k_I$, as shown in \cref{eq:voltage-and-current-mapping}.
\begin{subequations}\label{eq:voltage-and-current-mapping}
  \begin{align}
    V &= k_V x \\
    y &= \frac{I}{k_I} &\quad &\text{where $k_I = k_V k_G$}
  \end{align}
\end{subequations}
where $k_G$ is the conductance scaling factor typically made equal to $\dfrac{G_\mathrm{on} - G_\mathrm{off}}{\max|\matr{w}|}$ and $k_V$ is determined before training.

To encode both positive and negative weights, \emph{pairs} of conductances are employed.
Conductances $G_+$ and $G_-$ are introduced into ``positive'' and ``negative'' bit lines of crossbar arrays, where the output currents of the latter are subtracted from the output currents of the former; this is referred to as \newconcept{differential pair architecture}.
Each weight is typically made proportional to the difference of $G_+$ and $G_-$ (with $k_G$ acting as the constant of proportionality), which enables to encode any real number within a finite interval.
However, infinite conductance combinations will produce the same difference~\cite{KiMa2021}, thus the network designer may have to make an arbitrary choice of how to perform this mapping.
For example, to encode weights $w \in \matr{w}$, the two conductances may be picked symmetrically around the average value~\cite{KiMa2021}, as shown in \cref{eq:mapping-symmetric-G}.
\begin{equation}\label{eq:mapping-symmetric-G}
  G_\pm = G_\mathrm{avg} \pm \frac{k_G w}{2}
\end{equation}
where $G_\mathrm{avg} \equiv \dfrac{G_\mathrm{off} + G_\mathrm{on}}{2}$.

Although there might be advantages to using the mapping scheme like the one in \cref{eq:mapping-symmetric-G}\footnote{An argument could be made, for example, that such a mapping scheme would produce very few conductances with values near $G_\mathrm{off}$ or $G_\mathrm{on}$, which may be more difficult to achieve in real devices.}, the choice of mapping could be \emph{explicitly} tied to certain objectives.
For example,~\cite{LiHu2018} points out that differential pair architecture with aware mapping can be advantageous for mitigating the effects of stuck devices.
Alternatively, a mapping scheme that optimizes some metric (like power consumption) may be employed.
Indeed, such a scheme is used throughout this work for mapping the weights of conventionally trained \glspl{ANN} onto conductances---we minimize power consumption by ensuring that at least one device in the pair is set to $G_\mathrm{off}$, as demonstrated in \cref{eq:mapping-low-G}.
\begin{equation}\label{eq:mapping-low-G}
  \begin{aligned}
    G_+ &= G_\mathrm{off} + \max\{0, k_G w\} \\
    G_- &= G_\mathrm{off} - \min\{0, k_G w\}
  \end{aligned}
\end{equation}

However, choosing the optimal scheme manually is a low-level approach that requires understanding the physical characteristics of \glspl{MNN}.
Thus, even if training is done \textit{ex situ}, network designer has to make choices about not only the conventional training hyperparameters (like learning rate), but also how the system will be implemented physically because that will affect \gls{MNN} performance.
We see this as an additional obstacle to making memristive implementations of \glspl{ANN} feasible in the real world.

\subsubsection{Our approach---double weights}

In this work, instead of tweaking the mapping function, we decided to change the characteristics of the weights that are being trained.
Specifically, we train \emph{two} sets of \emph{nonnegative} weights, $w_{ij}^+ \in \matr{w_+} \in \mathbb{R}_{\geq 0}^{M \times N}$ and $w_{ij}^- \in \matr{w_-} \in \mathbb{R}_{\geq 0}^{M \times N}$, which we refer to as \newconcept{double weights}; similar approach has been explored in~\cite{KePa2020}.
We map double weights onto conductances in the aforementioned ``positive'' and ``negative'' bit lines, respectively.
Although all the weight parameters are nonnegative, the negative contribution of $i\textsuperscript{th}$ input on $j\textsuperscript{th}$ output can still be encoded because of the differential pair architecture in the \emph{physical} system.
Only the nonlinearity-aware \textit{ex-situ} training function in \cref{eq:synapse-nonlinearity-aware} has to be adjusted leading to the form in \cref{eq:synapse-nonlinearity-aware-double-weights}.

\begin{equation}\label{eq:synapse-nonlinearity-aware-double-weights}
  y_j = f{\left(\sum_{i = 1}^M {g \left(x_i, w_{ij}^+ \right) - g \left(x_i, w_{ij}^- \right)} \right)}
\end{equation}

The adoption of double weights allows to relate \emph{every} weight in $\matr{w} \equiv \begin{bmatrix}\matr{w_+} & \matr{w_-}\end{bmatrix}$ to the corresponding conductance in the \emph{same} way, i.e.\ $w_\pm \in \interval{0}{\max(\matr{w})}$ are linearly mapped onto $G_\pm \in \interval{G_\mathrm{off}}{G_\mathrm{on}}$, as shown in \cref{eq:mapping-double-w-G}.

\begin{equation} \label{eq:mapping-double-w-G}
  G_\pm = k_G w_\pm + G_\mathrm{off}
\end{equation}

A clear advantage of our approach is that double weights allow for more direct optimization of \gls{MNN} behavior.
Exposing raw device characteristics---i.e.\ conductances---to the training algorithm, enables it to select the combination that has both the optimal performance (as defined by some metric like loss) and high robustness.
For example, if a certain nonideality manifests itself to a greater degree at low conductances, the training algorithm would be able to push double weight pairs (and by extension, conductances) toward higher values.
Because of the differential pair architecture, setting $G_+$ and $G_-$ to \SI{1.0}{\milli\siemens} and \SI{2.0}{\milli\siemens}, respectively, will---at least in the case of linearity-preserving nonidealities---have the same effect as setting them to \SI{3.0}{\milli\siemens} and \SI{4.0}{\milli\siemens}, respectively.
Therefore, with double weights, the training algorithm should be able to choose conductance combinations that minimize the negative influence of nonidealities.

Additionally, double weights allow regularization to act as a high-level tool for controlling the importance of power consumption.
We propose training with the $\ell_1$ sparsification regularizer~\cite{HaPo2015}, which can not only improve training\footnote{This may manifest itself as avoiding overfitting, for example.}, but also promote lower conductances because they are linearly related to weight parameters, as demonstrated in \cref{eq:mapping-double-w-G}.
During backward propagation, the regularizer influences training loss, inducing conductance pairs to descend toward $G_\mathrm{off}$.
Instead of manually tweaking the mapping function, network designer can decide to what extent energy efficiency should be prioritized by simply adjusting, say, regularization factor in $\ell_1$ regularization.
This can be incorporated into typical hyperparameter tuning process that is performed before deploying \glspl{ANN} in practise.

\subsection{Modified Validation}\label{subsubsec:modifed-validation}

We also propose a modified model validation scheme more fit for \glspl{MNN}.
To determine when to stop the training (or which version of the network to use when the training takes a predetermined set of epochs), a validation dataset is commonly employed---a metric (like error or loss) is computed for this dataset at certain epochs and is used for picking the optimal version of the network~\cite{ShTa2010}.
However, many of memristor nonidealities are at least partly stochastic in nature, thus, say, validation accuracy at any given epoch may not be entirely representative of the model's quality purely due to random chance.
Because of this, we suggest computing the validation metric multiple times and using an aggregated value for higher reliability.
Choices can be made about
\begin{itemize}
  \item aggregate value that is used
  \item how frequently validation is performed
  \item how many times validation metric is computed at each checkpoint
\end{itemize}
In the simulations of this work, we computed validation error $20$ times every $20$ epochs and saved the model whenever the \emph{median} validation error decreased.

\subsection{Nonidealities}\label{subsec:nonidealities}

In this work, we explore a wide range of memristor nonidealities and utilize experimental data.
We use two different memristor technologies---\ce{SiO_x}- and \ce{Ta}/\ce{HfO2}-based \gls{RRAM}.
More details on the two technologies can be found in \cref{sec:experimental-section} and our previous publications~\cite{MeJo2019, JoFr2020}.

\subsubsection{\IV\ nonlinearity}

One of the most common ways to characterise deviations from ohmic behavior in memristive devices has been by considering two points on an \IV\ curve~\cite{SuLi2018, FlBe2013}.
For example, one may define nonlinearity at voltage $V_\mathrm{ref}$ as the ratio of the conductance at that voltage to the conductance at half that voltage~\cite{SuLi2018}, as shown in \cref{eq:nonlinearity-definition}.
Nonlinearity of $1$ can then be characterized as indicative of ohmic behavior; similarly, any deviations from that value indicate \IV\ nonlinearity.
This metric can be useful in describing non-ohmic behavior at different voltages but it is more challenging to utilize it for modeling purposes.
\begin{equation}\label{eq:nonlinearity-definition}
  \begin{aligned}
    \text{nonlinearity} &\equiv \frac{G(V_\mathrm{ref})}{G(V_\mathrm{ref}/2)} \\ \text{where } G(V) &\equiv \frac{\text{current at voltage $V$}}{\text{voltage $V$}}
  \end{aligned}
\end{equation}

\begin{figure*}[h!]
  \begin{center}
    \includegraphics{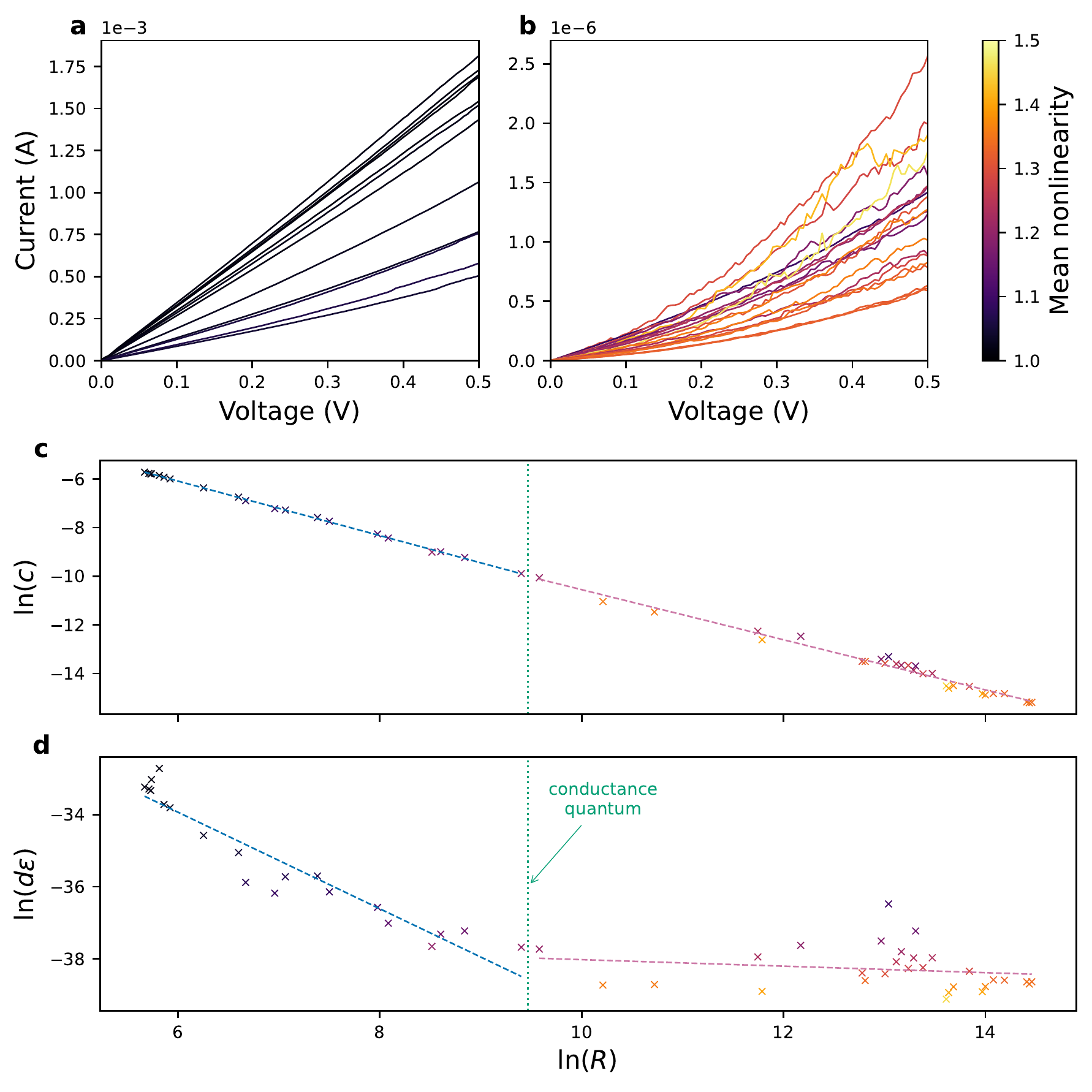}
    {\phantomsubcaption\label{fig:SiO_x:a}}
    {\phantomsubcaption\label{fig:SiO_x:b}}
    {\phantomsubcaption\label{fig:SiO_x:c}}
    {\phantomsubcaption\label{fig:SiO_x:d}}
  \end{center}
  \mycaption{\ce{SiO_x} data utilized in this work}{%
    \IV\ sweeps of a \ce{SiO_x} device are shown for \subf{a}) a subset of the low-resistance region (with resistance at \SI{0.1}{\volt} ranging from \SI{289.8}{\ohm} to \SI{1169}{\ohm}), and \subf{b}) a subset of the high-resistance region (with resistance at \SI{0.1}{\volt} ranging from \SI{445.2}{\kilo\ohm} to \SI{1.905}{\mega\ohm}).
    For all curves, only the range of voltages from \SI{0.0}{\volt} to \SI{0.5}{\volt} was considered.
    Nonlinearity was computed by dividing the conductance at voltage $V$ by the conductance at voltage $0.5V$.
    Poole-Frenkel fits in low- (left) and high-resistance (right) regions for \subf{c}) $c$ and \subf{d}) $d \varepsilon$.
    In both panels, to ensure dimensionless inputs to logarithms, $R$, $c$, and $d \varepsilon$ are the amounts of the corresponding quantities in SI units, e.g.\ $R$ is is not the resistance, but rather the \emph{number} of ohms of resistance measured at \SI{0.1}{\volt}.
    Marker colors represent the mean nonlinearity of the \emph{fits}, rather than the experimental \IV\ curves.
  }\label{fig:SiO_x}
\end{figure*}

In this work, we utilized silicon oxide devices to investigate the effects of current-voltage nonlinearity.
\ce{SiO_x} devices can undergo resistance switching---typical \IV\ switching curve is shown in \crefExt{supp-fig:SiO_x-switching} in the Supporting Information, while a more detailed analysis of resistance switching performance can be found in our previous study~\cite{MeMu2017}.
For the purposes of this work, to achieve a wide range of resistance states and to analyze \IV\ nonlinearity, incremental positive sweeps were used to gradually reset the device from the low-resistance state to the high-resistance state.
\IV\ curves of two subsets of all achieved states are shown in \cref{fig:SiO_x:a,fig:SiO_x:b}.
Low-resistance discrete states (in \cref{fig:SiO_x:a}) exhibit more linear behaviour and experience little variability in nonlinearity.
On the other hand, high-resistance states (in \cref{fig:SiO_x:b}) are more nonlinear and the nonlinearity is less predictable.

Multiple conduction mechanisms have been proposed for modeling \IV\ behavior of memristors.
The models may incorporate principles like Fowler-Nordheim tunneling, thermionic emission, and Poole-Frenkel emission~\cite{SzLi2006}.
Data in \cref{fig:SiO_x:a,fig:SiO_x:b} were best fit using Poole-Frenkel model in \cref{eq:poole-frenkel}.
\begin{equation}\label{eq:poole-frenkel}
  I = c V \exp \left( \frac{2e}{k_\mathrm{B} T} \sqrt{\frac{eV}{4 \pi d \varepsilon}} \right)
\end{equation}
where $I$ is the current, $V$ is the voltage, $c$ is a constant (which has units of conductance), $T$ is the temperature, $d$ is assumed to be oxide thickness or effective thickness for partially oxidized filament, and $\varepsilon$ is the permittivity.

$T$ is room temperature, thus the following parameters were fit: $c$ and the product of $d$ and $\varepsilon$.
Different sets of fits were produced for above and below the conductance quantum $G_0 = 2e^2/h$ because states below this value experienced different trends and/or amounts of variability.
This is not surprising as memristive devices have been reported to exhibit different behavior in different resistance states and that is often tied to the conductance quantum~\cite{YiSa2016, MeBu2015}.

The fits for the two sets of parameters are shown in \cref{fig:SiO_x:c,fig:SiO_x:d}.
Slopes close to $-1$ and intercepts close to $0$ in \cref{fig:SiO_x:c} demonstrate that $c$ does indeed act similarly to conductance, i.e.\ reciprocal of resistance.
In high-resistance states on the right, however, the variability is noticeably higher.
In \cref{fig:SiO_x:d}, product $d \varepsilon$ is indicative of the extent to which a given curve is nonlinear---the smaller this product, the more nonlinear the \IV\ curve is.
For the low-resistance states, $d \varepsilon$ decreases with increasing resistance; however, for the first few states, this product is large---\cref{eq:poole-frenkel} can approximate linear behavior (ohmic conduction); however, the values of $d \varepsilon$ are physically plausible only for the high resistance states (${<}G_0$). 
For those high-resistances states on the right side of \cref{fig:SiO_x:d}, the product is, on average, lower but there is no \emph{clear} trend \emph{between} those states.
It is also more difficult to predict $d \varepsilon$ from a given $R$ as the deviations from the linear fit are very large.
Thus, not only are highly resistive states more nonlinear, but there is also a significant level of variability.
This is also indicated in the colors of the curves in \cref{fig:SiO_x:b} where the last few states are shown---there is no obvious relationship between the resistance state and the color (which indicates nonlinearity).
During nonideality-aware training, we take into account not only the nonlinearity but also the variability of \IV\ curves.

The simulations were performed by considering two different resistance ranges from our experimental data and by subsequently using fits from \cref{fig:SiO_x:c,fig:SiO_x:d}.
Low-resistance group was constructed by interpolating between the lowest achieved resistance state of \SI{289.8}{\ohm} and five times that resistance.
Similarly, to ensure the same dynamic range, high-resistance group was constructed by interpolating between the highest achieved resistance state of \SI{1.905}{\mega\ohm} and one fifth that resistance\footnote{The curves of resistance states falling in either of these two intervals are the ones depicted in \cref{fig:SiO_x:a,fig:SiO_x:b}.}.
As hinted earlier, it is na\"{\i}ve to simply use the aforementioned linear fits without taking into account significant deviations.
In fact, the presence of uncertainty in the model is one of our main goals because we wish to demonstrate that nonideality-aware training can adapt not only to deviations from linear behavior but also to stochastic behavior.
This makes nonideality-aware training approach generalizable because it does not require \emph{exact} knowledge of device behavior---it improves the performance even when different hardware is used as will be demonstrated in \cref{sec:results-agnosticism}.

During simulations involving nonlinear \IV\ behavior, the output current of any given device was determined in the following way:
\begin{enumerate}
  \item $c$ and $d \varepsilon$ were interpolated from the fits in \cref{fig:SiO_x:c,fig:SiO_x:d} using resistance (parameter) $R$
  \item $c$ and $d \varepsilon$ were disturbed using multivariate normal distribution with the covariance matrix determined using the residuals of the fits
  \item current $I$ was computed using \cref{eq:poole-frenkel}
\end{enumerate}
We note that such treatment, where only the linear relationship between the two sets of residuals is considered, likely \emph{overestimates}, rather than underestimates, the amount of uncertainty for a given device.
Additional information on heteroscedasticity, the correlation between the residuals of the two sets of parameters, and the justification for using normal distribution can be found in \cref{subsec:statistical-analysis}.

Linearity-nonpreserving nonidealities like \IV\ nonlinearity cannot be simulated using conventional noise injection methods, which simply disturb the conductance values.
Instead, a forward propagation function must be defined reflecting the nonlinear relationship between inputs and outputs.
One can express the procedure described earlier in the form of the aforementioned function $g$ representing nonlinear behavior; this can be done by combining \cref{eq:voltage-and-current-mapping,eq:mapping-double-w-G,eq:poole-frenkel} leading to the form in \cref{eq:g-iv-nonlinearity}, which we implement using TensorFlow.
\begin{equation}\label{eq:g-iv-nonlinearity}
  \begin{aligned}
    g \left( x, w_\pm \right) &= \frac{1}{k_G} c x \exp \left( \frac{2e}{k_\mathrm{B} T} \sqrt{\frac{e k_V x}{4 \pi d \varepsilon}} \right) \\
    \text{where } \begin{bmatrix}
      c \\
      d \varepsilon
    \end{bmatrix} &= \exp\left( \ln\left( \frac{1}{k_G w_\pm + G_\mathrm{off}} \right) \matr{m} + \matr{b} + \matr{E} \right) \\
    \text{and } \matr{E} &\sim \mathcal{N}_2(\matr{0}, \matr{\Sigma})
  \end{aligned}
\end{equation}
where $\matr{m}$ and $\matr{b}$ are slopes and intercepts, respectively, of the corresponding fits in \cref{fig:SiO_x:c,fig:SiO_x:d}, $\matr{\Sigma}$ is the covariance matrix of the residuals and all inputs to logarithms or exponents are the amounts of quantities in SI units.

\subsubsection{Stuck devices}

\begin{figure*}[b]
  \begin{center}
    \includegraphics{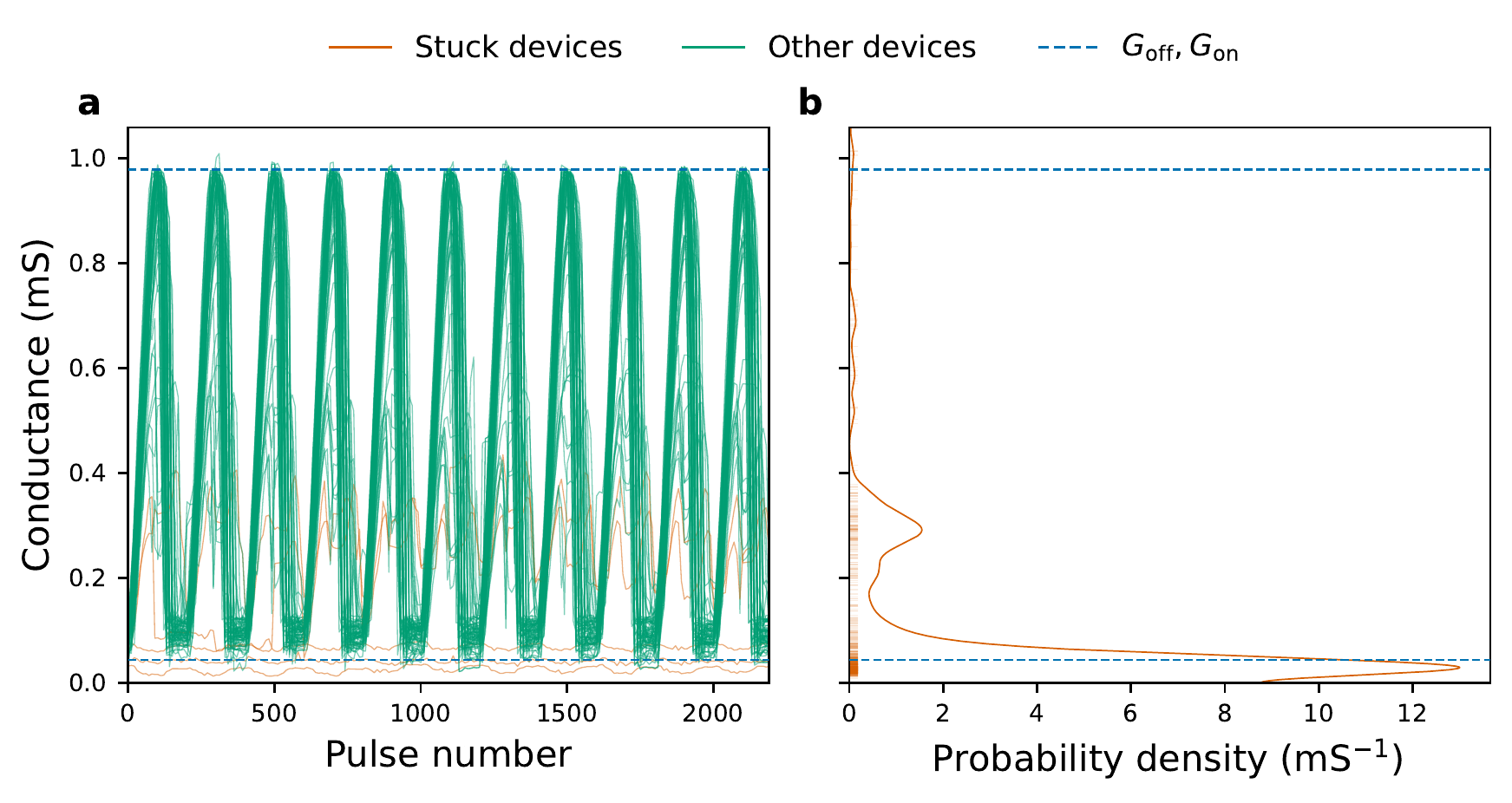}
    {\phantomsubcaption\label{fig:Ta-HfO2:a}}
    {\phantomsubcaption\label{fig:Ta-HfO2:b}}
  \end{center}
  \mycaption{Experimental \ce{HfO2} data utilized in this work}{%
    \subf{a}) $11$ potentiation and depression cycles of $82$ \ce{Ta}/\ce{HfO2} devices---data are shown for \SI{1}{\percent} of the devices in a $128 \times 64$ crossbar array.
    Any device whose maximum conductance range was less than half the median range was classified as ``stuck''.
    The conductance of \ce{Ta}/\ce{HfO2} crossbar devices was modulated by controlling the voltage pulses applied to the gate of the selector transistor; more details are provided in \cref{sec:experiments-Ta-HfO2} and~\cite{WaLi2019}.
    \subf{b}) Mean conductances of all stuck devices together with their estimated probability density function constructed using Gaussian kernel density estimation.
  }\label{fig:Ta-HfO2}
\end{figure*}

Additionally, we explore linearity-preserving nonidealities, which can be simulated using noise injection into the conductances.
One of such nonidealities is devices stuck in a particular state, which is a very common issue in memristors.
The effect of stuck devices can be explored in isolation, but it also easily lends itself to being simulated along other nonidealities thus allowing to investigate the effectiveness of nonideality-aware training in more complex scenarios.
In the modeling of this nonideality, we consider both real experimental data (where we draw the state in which devices may get stuck from a probability distribution) and a simplified model (where we assume devices can get stuck in only one state).

Data from $128 \times 64$ \ce{Ta}/\ce{HfO2} memristor crossbar array were analyzed for the purposes of modeling stuck devices' behavior.
\Cref{fig:Ta-HfO2:a} shows $11$ potentiating and depression cycles (each consisting of a \num{100} voltage pulses) for a fraction of the devices.
By considering the minimum and maximum conductance values achieved by each of the \num{8192} devices over \num{11} cycles, $G_\mathrm{off}$ and $G_\mathrm{on}$ were chosen as the median of these minimum and maximum values, respectively.
Devices whose maximum range was less than half the median range (where median range had been defined as $G_\mathrm{on} - G_\mathrm{off}$) were classified as ``stuck''.

Stuck devices were simulated using a probabilistic model.
Using the aforementioned stuck devices' definition, \SI{10.1}{\percent} of the devices were classified as such.
To simplify modeling, these devices were simulated to be \emph{fully} stuck, meaning their conductance could not be changed even by a small amount\footnote{Similar to uncertainty in \ce{SiO_x} devices, this assumption \emph{overestimates} the effect of stuck \ce{HfO2} devices, not underestimates it.}.
The average values of all stuck devices are denoted by markers on the $y$ axis of \cref{fig:Ta-HfO2:b}.
The probability density function of these average values was constructed using kernel density estimation with truncated normal distributions.
Scott's rule~\cite{Sc1992} was used for bandwidth estimation, while mirror reflections of the distributions were employed to correct for bias at the \SI{0}{\siemens} clipping boundary\footnote{To ensure numerical stability, mirror reflections were only included if the area under the curve of the underlying normal distribution below \SI{0}{\siemens} was more than $10^{-8}$.}.
When simulating the effect of the nonideality, each device could be set to any conductance state between $G_\mathrm{off}$ and $G_\mathrm{on}$; however, every device also had a \SI{10.1}{\percent} probability of getting stuck.
When a device was classified as stuck, its conductance would be drawn from the probability distribution constructed using kernel density estimation in \cref{fig:Ta-HfO2:b}.

\begin{figure*}[b]
  \begin{center}
    \includegraphics{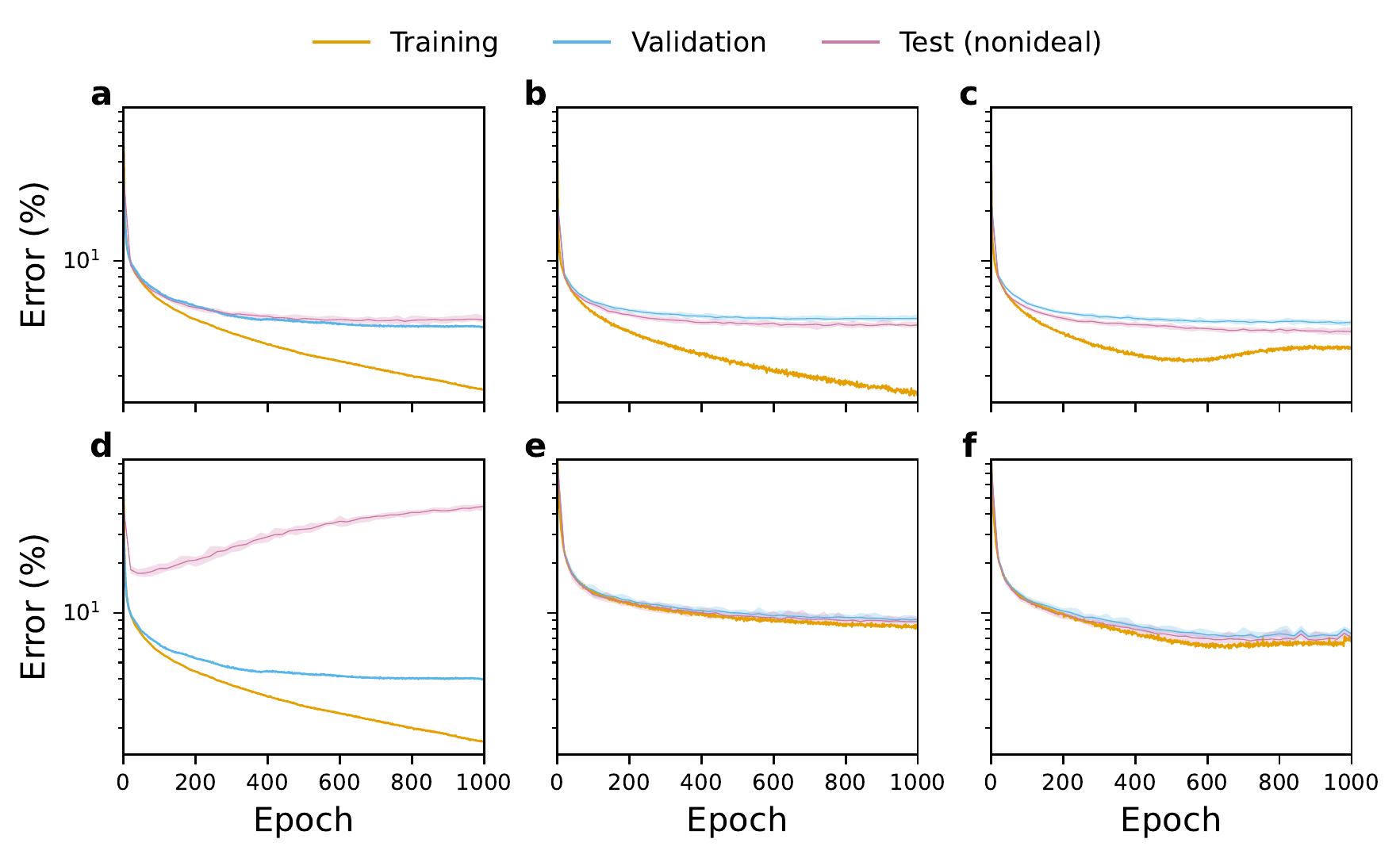}
    {\phantomsubcaption\label{fig:iv-nonlinearity-training:a}}
    {\phantomsubcaption\label{fig:iv-nonlinearity-training:b}}
    {\phantomsubcaption\label{fig:iv-nonlinearity-training:c}}
    {\phantomsubcaption\label{fig:iv-nonlinearity-training:d}}
    {\phantomsubcaption\label{fig:iv-nonlinearity-training:e}}
    {\phantomsubcaption\label{fig:iv-nonlinearity-training:f}}
  \end{center}
  \mycaption{Training results for standard and nonideality-aware schemes when exposed to \IV\ nonlinearities}{%
    \subf{a}--\subf{c}) low \IV\ nonlinearity (data from \cref{fig:SiO_x:a}), \subf{d}--\subf{f}) high \IV\ nonlinearity (data from \cref{fig:SiO_x:b}); \subf{a},\,\subf{d}) standard training (same training and validation curves), \subf{b},\,\subf{e}) nonideality-aware training, \subf{c},\,\subf{f}) nonideality-aware training with regularization.
    The panels show curves for one of five sets of trained networks.
    Networks were trained on MNIST dataset.
    Where error was computed multiple times, the curves show the median value, as well as the region bounded by the minimum and maximum values.
  }\label{fig:iv-nonlinearity-training}
\end{figure*}

Additionally, a simple model of devices getting stuck in $G_\mathrm{off}$ or $G_\mathrm{on}$ was considered.
This allowed to combine nonidealities together, specifically with \ce{SiO_x} \IV\ nonlinearities.
For both states, devices were simulated to have \SI{5}{\percent} probability of getting stuck.
As with experimental stuck devices' model, this was simulated as noise injection into the conductances.
Both models of stuck device behavior were used to test the robustness of nonideality-aware training.

\subsubsection{\Glsentrylong{D2D} variability}

We also consider \gls{D2D} variability arising from inaccuracies during device programming.
During mapping onto conductances, one may end up with different values than intended; in some memristors, these (resistance) deviations are modeled using lognormal distribution~\cite{KiYa2016}.
As with stuck devices, this can be incorporated into training by disturbing the values in each iteration---in this case, by drawing from lognormal distribution.
We use this nonideality mostly to explore
\begin{enumerate}
  \item double weights (see \cref{sec:results-weight-implementation})
  \item the effects of incredibly stochastic nonidealities (see \cref{sec:results-agnosticism})
\end{enumerate}

For the lognormal modeling, we linearly interpolate the standard deviation of the natural logarithm of resistances from the following values meant to represent different device behaviors:
\begin{itemize}
  \item $0.25$ for $R_\mathrm{off}$ and $R_\mathrm{on}$ (more uniform \gls{D2D} variability)
  \item $0.5$ for $R_\mathrm{off}$ and $0.05$ for $R_\mathrm{on}$ (less uniform \gls{D2D} variability)
  \item $0.5$ for $R_\mathrm{off}$ and $R_\mathrm{on}$ (high-magnitude \gls{D2D} variability)
\end{itemize}

\section{Results and Discussion}

\subsection{Nature of Training}\label{subsec:nature-of-training}

\Cref{fig:iv-nonlinearity-training} contains training curves for \glspl{MNN} trained on MNIST dataset and exposed to \IV\ nonlinearities.
Although not used to affect any aspect of the training process, error curves for the test subset\footnote{Test set errors at training checkpoints were computed in the same way as memristive validation curves, see \cref{subsubsec:modifed-validation}.} are included as well.
Testing on nonideal configuration during training can help understand the differences between standard and nonideality-aware training.

\Cref{fig:iv-nonlinearity-training:a,fig:iv-nonlinearity-training:b,fig:iv-nonlinearity-training:c} explore the effect of low \IV\ nonlinearity.
In \cref{fig:iv-nonlinearity-training:a}, the validation curve (which is computed assuming digital implementation of the \gls{ANN}) is closely coupled with the test curve (which assumes that nonideal effects \emph{are} present)---this suggests negligible effect of low \IV\ nonlinearity.
Consequently, nonideality-aware training produces similar results on the test set both without (\cref{fig:iv-nonlinearity-training:b}) and with (\cref{fig:iv-nonlinearity-training:c}) regularization.

\Cref{fig:iv-nonlinearity-training:d,fig:iv-nonlinearity-training:e,fig:iv-nonlinearity-training:f} explore the effect of high \IV\ nonlinearity.
In \cref{fig:iv-nonlinearity-training:d}, where the results of standard training are presented, we notice that validation and test curves are detached from one another.
Not only that, but the global minimum of the (nonideal) test curve occurs very early in the training, while the (ideal) validation error keeps decreasing.
This indicates that without taking nonidealities into account during training, a highly suboptimal version of the \gls{ANN} may be chosen for inference stage with nonidealities.
Nonideality-aware training without (\cref{fig:iv-nonlinearity-training:e}) and with (\cref{fig:iv-nonlinearity-training:f}) regularization is much more effective---the validation and test curves are closely coupled together and the test error decreases to lower values.

\subsection{Performance Improvement}\label{subsec:performance-improvement}

\begin{figure}[t]
  \begin{center}
    \includegraphics{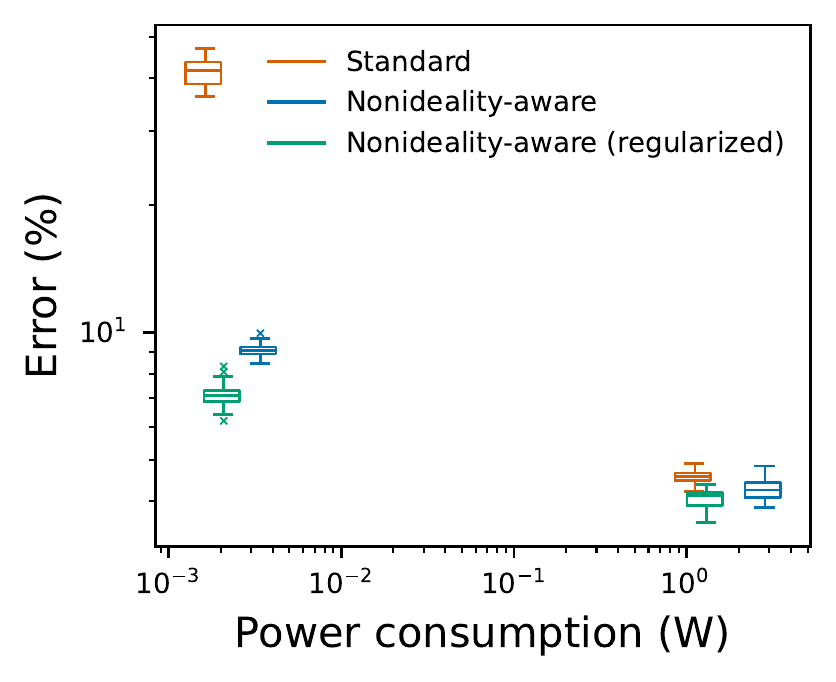}
  \end{center}
  \mycaption{Inference results of using nonideality-aware training to deal with \IV\ nonlinearities}{%
    The three box plots on the right refer to \glsentrylongpl{MNN} that used data from \cref{fig:SiO_x:a}, and the three on the left---to \glsentrylongpl{MNN} that used data from \cref{fig:SiO_x:b}.
    Networks were trained on MNIST dataset.
    The variability in power consumption for the data points of each group is small thus the \emph{average} power consumption is used for the horizontal position of each box plot.
  }\label{fig:iv-nonlinearity-inference}
\end{figure}

Inference results for \IV\ nonlinearity simulations are summarized in \cref{fig:iv-nonlinearity-inference}.
Because the simulated memristors are nonlinear, power consumption was computed using $P = IV$, instead of $P = I^2 R$, for each of the individual devices.
Although, apart from crossbar arrays, \gls{MNN} implementations require additional circuitry,~\cite{ChNa2019} suggest power consumption due to passive elements \emph{dominates} in the \si{\micro\siemens} range and above (i.e.\ the conductance range of the device investigated in this work); only at lower conductances does the relative impact of other energy components become significant.

Box plots representing low-resistance (and, by extension, low-nonlinearity) devices are presented on the right side of \cref{fig:iv-nonlinearity-inference}.
With such devices, nonideality-aware training without regularization achieves median error of \SI{4.3}{\percent}.
This is close to the error---\SI{4.6}{\percent}---of (digital) \glspl{ANN} of the same size trained using standard procedure; the small difference is indicative of the relatively small effect of low \IV\ nonlinearity (and its uncertainty).
However, regularization not only decreases the power consumption by more than half, but also helps achieve median error rate of \SI{4.1}{\percent}.

As shown on the left side of \cref{fig:iv-nonlinearity-inference}, \glspl{MNN} implemented using high-resistance devices achieve almost three orders of magnitude lower power consumption.
However, \glspl{MNN} trained using the standard procedure have median error of \SI{41.6}{\percent}, which would be unacceptable in most scenarios.
Fortunately, adjusted training results in much lower error rate, while maintaining low power consumption (compared to low-nonlinearity devices).
In the non-regularized case, the median error is \SI{9.1}{\percent}, and in regularized \glspl{MNN} it is \SI{7.1}{\percent}.
Thus, nonideality-aware training makes it feasible to use orders of magnitude more power-efficient high-resistance devices while maintaining error rates similar to those achieved with low-resistance devices.

One may also compare estimated absolute energy efficiency in both cases.
By assuming the values used in~\cite{YaWu2020}---read pulses of \SI{50}{\nano\second}, and \num{2} operations per synaptic weight (multiplication and accumulation)---one can calculate energy efficiency in \si{OP\second^{-1}\watt^{-1}} using \cref{eq:energy-efficiency}.
\begin{equation}\label{eq:energy-efficiency}
  \text{energy efficiency} = \frac{2 \times n}{50 \times 10^{-9} \times P_\mathrm{avg}}
\end{equation}
where $n$ is the number of synaptic weights and $P_\mathrm{avg}$ is the average power consumption.

\begin{figure*}[b]
  \begin{center}
    \includegraphics{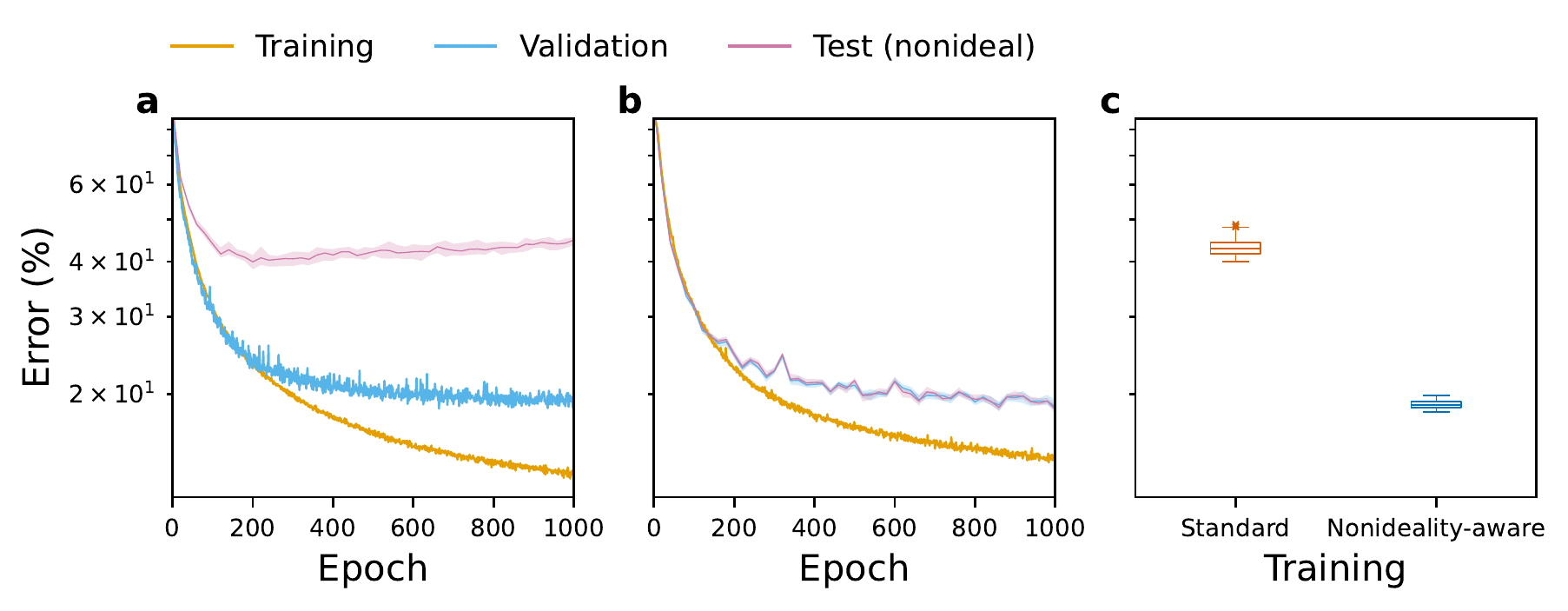}
    {\phantomsubcaption\label{fig:iv-nonlinearity-cnn:a}}
    {\phantomsubcaption\label{fig:iv-nonlinearity-cnn:b}}
    {\phantomsubcaption\label{fig:iv-nonlinearity-cnn:c}}
  \end{center}
  \mycaption{Results for standard and nonideality-aware schemes when employed by \glsentrylongpl{CNN}}{%
    \subf{a}) standard training, \subf{b}) nonideality-aware training, \subf{c}) inference error comparing the approaches in \subf{a} and \subf{b}.
    Networks were trained on CIFAR-10 dataset and their fully connected layers were exposed to high \IV\ nonlinearity (data from \cref{fig:SiO_x:b}) during inference.
    Panels~\subf{a} and~\subf{b} show curves for one of five sets of trained networks.
  }\label{fig:iv-nonlinearity-cnn}
\end{figure*}

Using these assumptions, standard training using low-resistance devices achieves energy efficiency of \SI{0.715}{\tera OP\second^{-1}\watt^{-1}}, while nonideality-aware training using high-resistance devices achieves energy efficiency of \SI{234}{\tera OP\second^{-1}\watt^{-1}} in the nonregularized case and of \SI{381}{\tera OP\second^{-1}\watt^{-1}} in the regularized case.
As explained earlier, these estimates incorporate power consumption only on crossbar arrays.

\subsection{More Complex Architectures and Datasets}

To understand how well nonideality-aware training performs on more complex tasks, we employed CIFAR-10 dataset.
For this, we trained \glspl{CNN} assuming that their convolutional layers would be implemented digitally, and their fully connected layers---using memristive crossbar arrays suffering from high \IV\ nonlinearity.
Standard training is explored in \cref{fig:iv-nonlinearity-cnn:a}, and nonideality-aware training---in \cref{fig:iv-nonlinearity-cnn:b}.
As with \glspl{MNN} trained on MNIST, there is a much greater coupling between validation and test curves when nonidealities are taken into account.
As shown in \cref{fig:iv-nonlinearity-cnn:c}, nonideality-aware approach reduces the median inference error from \SI{43.0}{\percent} to \SI{18.9}{\percent}.

\subsection{Importance of Weight Implementation}\label{sec:results-weight-implementation}

\begin{figure*}[t!]
  \begin{center}
    \includegraphics{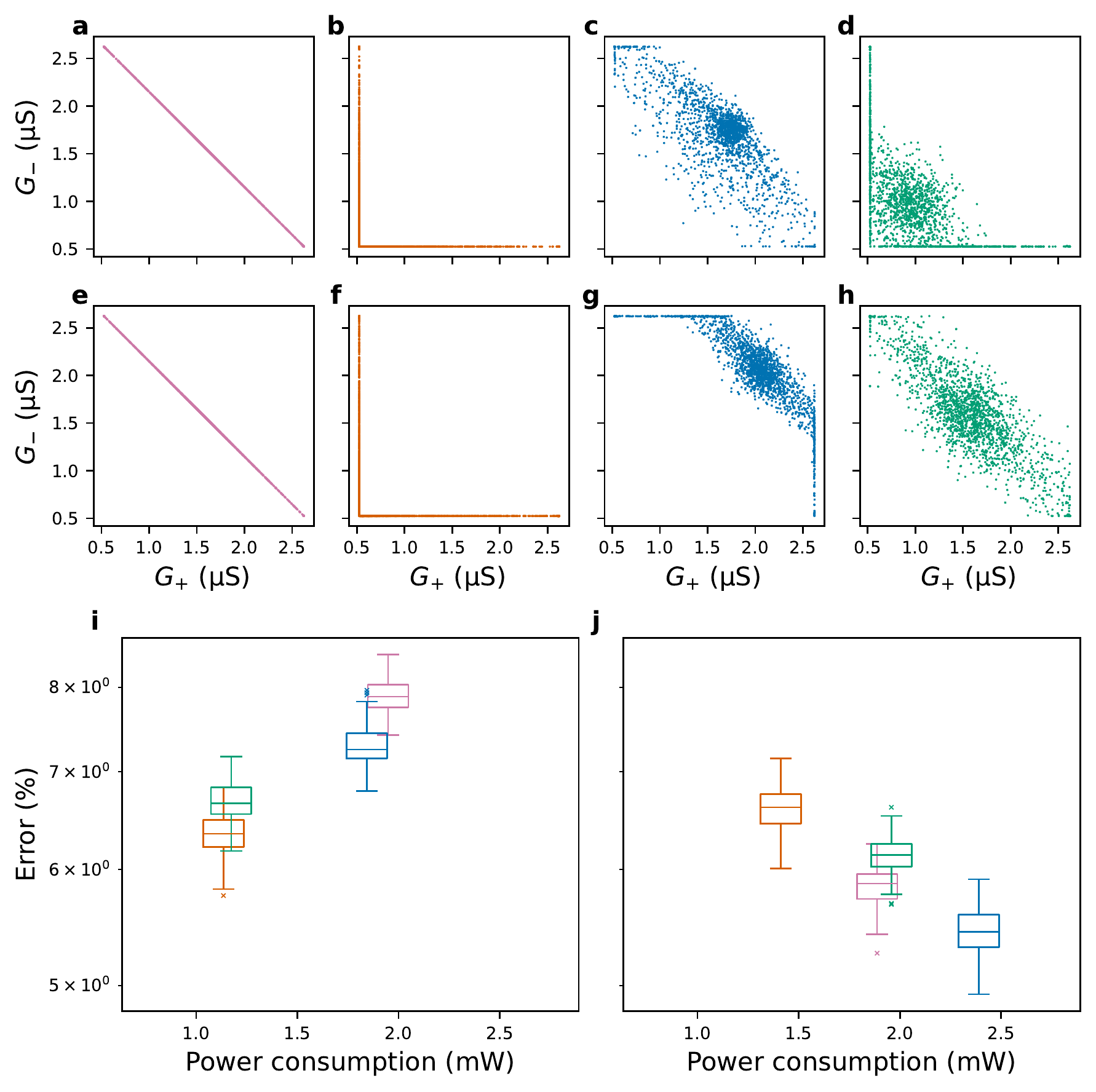}
    {\phantomsubcaption\label{fig:weight-implementation:a}}
    {\phantomsubcaption\label{fig:weight-implementation:b}}
    {\phantomsubcaption\label{fig:weight-implementation:c}}
    {\phantomsubcaption\label{fig:weight-implementation:d}}
    {\phantomsubcaption\label{fig:weight-implementation:e}}
    {\phantomsubcaption\label{fig:weight-implementation:f}}
    {\phantomsubcaption\label{fig:weight-implementation:g}}
    {\phantomsubcaption\label{fig:weight-implementation:h}}
    {\phantomsubcaption\label{fig:weight-implementation:i}}
    {\phantomsubcaption\label{fig:weight-implementation:j}}
  \end{center}
  \mycaption{Comparison of weight implementations}{%
    Conductance distributions in the first synaptic layer of a \glsentrylong{MNN} that deals with more uniform \glsentrylong{D2D} variability and utilizes \subf{a}) conventional weights mapped symmetrically around average conductance, \subf{b}) conventional weights mapped onto devices by preferring lowest total conductance, \subf{c}) double weights, \subf{d}) double weights with regularization; \subf{e}--\subf{h}) corresponding conductance distributions for a \glsentrylong{MNN} dealing with \emph{less} uniform \glsentrylong{D2D} variability.
    Scatter plots contain conductance data from \SI{10}{\percent} of the devices for one of five sets of networks.
    Inference error for different weight implementations of \glsentrylongpl{MNN} that deal with \subf{i}) more and \subf{j}) less uniform \glsentrylong{D2D} variability, where the box plot colors correspond to different weight implementations in \subf{a}--\subf{d} and \subf{e}--\subf{h}, respectively.
  }\label{fig:weight-implementation}
\end{figure*}

Double weights can be advantageous because they expose conductances to the training process in a more direct way.
To investigate the utility of this modified weight implementation, we utilized lognormal \gls{D2D} variability of two kinds:
\begin{enumerate}
  \item more uniform variability where the relative magnitude of deviations is the same throughout $\interval{G_\mathrm{off}}{G_\mathrm{on}}$
  \item less uniform variability where the relative magnitude of deviations is much greater near $G_\mathrm{off}$ compared to $G_\mathrm{on}$---similar to what is experienced in real devices when trying to program them~\cite{KiYa2016}
\end{enumerate}
This allowed to
\begin{enumerate}
  \item evaluate the performance of double weight implementation when the severity of nonideality does \emph{not} depend on the conductance value
  \item test whether double weight implementation would outperform standard weights when exposed to nonidealities whose severity depends on the conductance value
\end{enumerate}
Both standard weights with different mapping schemes and double weights without and with regularization are investigated in \cref{fig:weight-implementation}.

Firstly, we consider weight implementations in the context of \gls{D2D} variability with relatively high uniformity across the conductance range.
\Cref{fig:weight-implementation:a,fig:weight-implementation:b} show conductances resulting from mapping conventional weights using rules in \cref{eq:mapping-symmetric-G,eq:mapping-low-G}, respectively.
In both cases, the conductances are all mapped along either one or two line segments.
In contrast, \cref{fig:weight-implementation:c,fig:weight-implementation:d} show conductances obtained from double weight implementation and training without and with regularization, respectively.
In the nonregularized case (\cref{fig:weight-implementation:c}), the conductances are distributed mostly around the diagonal, though they are spread out, unlike in \cref{fig:weight-implementation:a}.
Regularization results in more data points in the bottom left corner of the diagram representing low conductance values, as shown in \cref{fig:weight-implementation:d}.

\Cref{fig:weight-implementation:e,fig:weight-implementation:f,fig:weight-implementation:g,fig:weight-implementation:h} demonstrate the utility of double weights.
Here, the \glspl{MNN} were trained to deal with less uniform \gls{D2D} variability where the disturbances are much greater at low conductance values compared to high conductance values.
As demonstrated in \cref{fig:weight-implementation:g}, the training naturally results in most of the pairs concentrated in the top right corner representing high conductance values.
Like before, regularization in \cref{fig:weight-implementation:h} results in conductances with lower values.

\Cref{fig:weight-implementation:i} shows the inference error for various weight implementations of \glspl{MNN} encountering more uniform \gls{D2D} variability.
Conventional weights with mapping rule in \cref{eq:mapping-low-G} result in the lowest median error of \SI{6.3}{\percent}; in comparison, double weights achieve median error of \SI{7.2}{\percent} without regularization and median error of \SI{6.7}{\percent} with regularization.
Conventional weights with mapping rule in \cref{eq:mapping-symmetric-G} achieve the highest median error of \SI{7.9}{\percent}, indicating the unpredictability of the performance of mapping methods.
Even so, it is important to point out that double weights do not achieve optimal performance in this case.
We hypothesize that in scenarios where the dependence of the severity of the nonideality on the conductance values is not strong, double weights might struggle to find optimal configuration---infinitely many pairs may result in the same behavior, making it a more computationally difficult problem.

\Cref{fig:weight-implementation:j} shows the inference error for equivalent weight implementations of \glspl{MNN} dealing with \emph{less} uniform \gls{D2D} variability.
In this case, the advantage of double weights is much more apparent---the median error in nonregularized case is \SI{5.4}{\percent} compared to \SI{5.9}{\percent} and \SI{6.6}{\percent} resulting from conventional weights with mapping rules in \cref{eq:mapping-symmetric-G,eq:mapping-low-G}, respectively.
When double weights are trained by employing regularization, they take on lower values thus decreasing power consumption but also increasing the error---in the case of this specific nonideality, there \emph{is} a tradeoff between energy efficiency and accuracy.
However, double weights together with regularization provide a straightforward way of specifying to what extent low power consumption should be prioritized at the expense of accuracy.

\subsection{Memristive Validation}\label{sec:results-memristive-validation}

As explained in \cref{subsubsec:modifed-validation}, many memristive nonidealities are nondeterministic, therefore it might be advantageous to compute an aggregate metric for use in validation.
During training, with each batch, we simulate nonidealities separately, e.g.\ parameters for \IV\ nonlinearity are drawn from a probability distribution or the exact devices that get stuck are picked randomly each time.
As a result, we believe that memristive validation can provide more reliable estimates of performance during training.
Although we hypothesize that \emph{in aggregate} this method will achieve only marginally better performance, it should help avoid choosing a highly suboptimal version of the weights, which might yield higher error in a \emph{small number} of cases.

Of course, memristive validation parameters may have to be optimized individually for each training configuration.
For example, in the memristive \gls{CNN} training in \cref{fig:iv-nonlinearity-cnn:b}, the variability of validation error is usually lower at \emph{any given checkpoint} than \emph{between} checkpoints.
In that case, one may increase the frequency of checkpoints by either decreasing the number of repeats at each ckeckpoint (and thus increasing uncertainty) or keeping it the same (and thus increasing computation time).
At the extreme---if one can afford additional training time---validation error may be computed every epoch multiple times.

\subsection{Nonideality Agnosticism}\label{sec:results-agnosticism}

Accurate modeling of nonidealities for nonideality-aware \textit{ex-situ} training is a significant challenge.
Firstly, the nature of nonidealities encountered in practice may be different than what was modeled for the purposes of training.
For example, the existence of \gls{D2D} variability in \ce{SiO_x} memristors means that the behavior of any individual device is not perfectly representative of the nature of other devices.
Therefore, to hedge against fitting the model to the behavior of any specific device, we assumed that Poole-Frenkel parameters are inferred using a linear fit (determined by a trend in the experimental data) \emph{and} disturbed by drawing random deviations from a probability distribution.
Even so, in different devices, these trends and amounts of deviations may be different.
Secondly, in the real world, one may encounter completely different types of nonidealities.
If the training takes into account the effects of only, say, \IV\ nonlinearities, \glspl{MNN} could still suffer from, for example, stuck devices when deployed.
Therefore, it is important to find out how robust the \glspl{MNN} employing nonideality-aware training are.

\begin{figure*}[b]
  \begin{center}
    \includegraphics{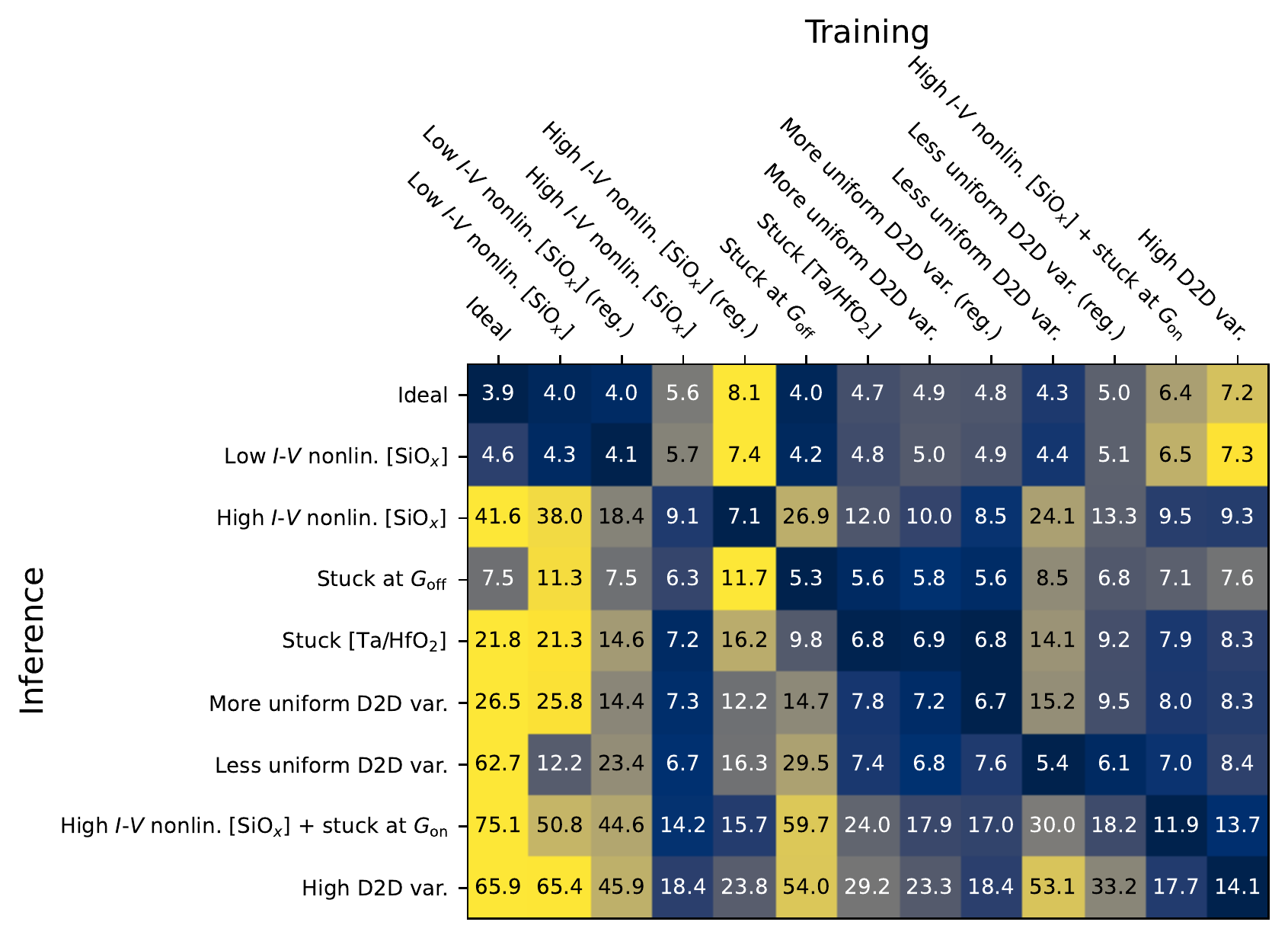}
  \end{center}
  \mycaption{Extent of nonideality agnosticism of nonideality-aware training}{%
    Median inference error in percent is shown for various training setups.
    Compared to standard (ideal) training, nonideality-aware approach usually results in lower error rate during inference, even if the nonideality encountered is different from the one the networks were exposed to during training.
    This is especially the case with more severe nonidealities.
    Networks were trained on MNIST dataset.
    Each row in the heatmap uses a separate instance of the logarithmically scaled colormap.
    Additional information on training and inference setups can be found in \crefExt{supp-tab:figure-7-setups} in the Supporting Information.
  }\label{fig:nonideality-agnosticism}
\end{figure*}

To investigate this, we utilized networks trained either by assuming no nonidealities or by being exposed to one of \num{8} different combinations of nonidealities.
In the case of both types of \IV\ nonlinearity and two types of \gls{D2D} variability, networks were additionally trained using regularization.
During inference, each group of networks was then exposed to the
\begin{itemize}
  \item setup that they were trained on
  \item setups of the other groups of networks
\end{itemize}
In total, this produced $(1 + 8 + 2 + 2) \times (1 + 8) = 117$ scenarios; median inference error for each is presented in the heatmap in \cref{fig:nonideality-agnosticism}.

The heatmap shows that, in most cases, the lowest error for a given nonideality during inference is achieved by the network that was exposed to it during training.
A few exceptions exist, including regularized networks that were exposed to more uniform \gls{D2D} variability during training---when exposed to high \IV\ nonlinearity during inference, they achieve lower error of \SI{8.5}{\percent} compared to the error of \SI{9.1}{\percent} of networks that were trained (without regularization) on this exact nonideality.
This may suggest that the nature of some nonidealities might often overlap and, when additional techniques like regularization are employed, the performance might be increased beyond even what can be achieved with the knowledge of a particular nonideality.

Importantly, \cref{fig:nonideality-agnosticism} demonstrates that nonideality-aware training makes networks robust to the effects a wide array of nonidealities.
Except for the ideal case and low-severity nonidealities (low \IV\ nonlinearity and devices stuck at $G_\mathrm{off}$), nonideality-aware training results in lower median error compared to conventional training for \emph{all} \gls{MNN} groups---even when the networks encounter different nonidealities during inference.
The heatmap also suggests that the effect of regularization on robustness is not always the same.
In the case of high \IV\ nonlinearity, regularization usually results in higher error (compared to nonregularized case) when encountering different nonidealities.
On the other hand, regularization in networks dealing with \gls{D2D} variability produces more robust behavior---when encountering different nonidealities, they usually achieve lower error.

\section{Conclusion}

In this work, design a novel nonideality-aware training scheme that can improve the performance of \glspl{MNN}.
We demonstrate the importance of taking nonideal device behavior into account during training---without that, training performance may not be indicative of how well the networks will perform during inference.
Importantly, we show the utility of nonideality-aware training in dealing with linearity-nonpreserving nonidealities, specifically \IV\ nonlinearity.
Our simulations show that the dichotomy between stable device behavior and power efficiency can become less relevant if the training stage is adjusted.
Indeed, if nonideality-aware training is used, high-nonlinearity devices may achieve similar error rate, while also having almost three orders of magnitude better energy efficiency of \SI{381}{\tera OP\second^{-1}\watt^{-1}} (with regularization) compared to \SI{0.715}{\tera OP\second^{-1}\watt^{-1}} with the conventional training scheme and low-resistance devices.

We additionally explore a number of adjacent factors that are worth considering when dealing with \gls{MNN} training.
For example, we explore the use of double weights as a way to control conductance values of \glspl{MNN} more directly trough methods like $\ell_1$ regularization.
We explore the need to consider how validation is performed during training when nonidealities are stochastic and thus the outputs of an \glsentryshort{MNN} are nondeterministic.
In contrast to many previous works, we also investigate how robust nonideality-aware training is---we find that, compared to conventional training, it can usually achieve much lower error when encountering nonidealities that it was not trained to deal with.
Besides, nonideality-aware training can be applied not just to one single type of nonideality---it can deal with multiple nonidealities at once, for example \IV\ nonlinearity \emph{and} stuck devices, as shown in \cref{fig:nonideality-agnosticism}.

Nonideality-aware training schemes are critical to making memristor-based \glspl{ANN} feasible.
Such schemes are key to ensuring low error rate and high energy efficiency.
Using experimental data and several novel optimization techniques, we demonstrate that our training design deals with a wide range of nonidealities, importantly linearity-nonpreserving nonidealities, which have not been addressed during \textit{ex-situ} training before.
Further, we demonstrate that high-resistance operational ranges can be used to reduce power consumption by almost three orders of magnitude without significant accuracy loss, despite the high level of associated nonidealities.

\section{Experimental Section}\label{sec:experimental-section}

\subsection{Experiments}

\subsubsection{\ce{SiO_x}}\label{sec:SiO_x-experiments}

\threesubsection{Fabrication}

\ce{SiO_x} \gls{RRAM} devices were fabricated on a \ce{Si} substrate with \SI{1}{\micro\metre} of thermal oxide on top.
A \SI{100}{\nano\metre} of \ce{Mo} was deposited on top of the \ce{Si}/\ce{SiO2} substrate that served as the bottom electrode.
The \ce{SiO_x} layer was sandwiched between the bottom and top electrodes and was deposited by reactive sputtering.
The top electrodes consisted of \SI{5}{\nano\metre} \ce{Ti} wetting layer followed by a \SI{100}{\nano\metre} of \ce{Au}; their pattern was defined using a shadow mask.
The device sizes ranged from $\SI{200}{\micro\metre} \times \SI{200}{\micro\metre}$ to $\SI{800}{\micro\metre} \times \SI{800}{\micro\metre}$.

\threesubsection{Characterization}

Electrical characterization of a $\SI{400}{\micro\metre} \times \SI{400}{\micro\metre}$ device was performed using Keithley 4200A-SCS\@.
The signals were applied to the top electrode (\ce{Au}), while the bottom electrode (\ce{Mo}) was connected to the ground.
The device required an initial electroforming step before stable resistive switching could be achieved.
The forming process was carried out by a negative voltage sweep, which stopped when the current had reached the limit of \SI{3}{\milli\ampere}.
Subsequently, \num{18} voltage sweeps were performed to guarantee proper device performance: the voltage was ramped from \SI{0.0}{\volt}, to \SI{\pm2.5}{\volt}, and back to \SI{0.0}{\volt} using a \SI{3}{\milli\ampere} current compliance.

After this, to achieve a wide range of resistances, incremental positive sweeps were applied to the sample, starting from \SI{0.5}{\volt} and increasing by \SI{0.05}{\volt} in each run.
This was being repeated until there was no further resistance change, i.e.\ the filament had returned to its initial (post-forming) state.
The obtained \IV\ curves are shown in \crefExt{supp-fig:all-SiO_x-IV-curves-full-range} in the Supporting Information, while a subset of curves utilized in this work are shown in \cref{fig:SiO_x:a,fig:SiO_x:b}.

\subsubsection{\ce{Ta}/\ce{HfO2}}\label{sec:experiments-Ta-HfO2}

\threesubsection{Fabrication}

\ce{Ta}/\ce{HfO2} 1T1R array contains NMOS transistors (with feature size of \SI{2}{\nano\metre}) and \ce{Pt}/\ce{HfO2}/\ce{Ta} \gls{RRAM} devices.
The bottom electrode was deposited by evaporating \SI{20}{\nano\metre} \ce{Pt} layer on top of a \SI{2}{\nano\metre} \ce{Ta} adhesive layer.
A \SI{5}{\nano\metre} \ce{HfO2} switching layer was deposited by atomic layer deposition using water and tetrakis(dimethylamido)hafnium as precursors at \SI{250}{\celsius}.
\SI{50}{\nano\metre} \ce{Ta} sputtered layer followed by \SI{10}{\nano\metre} \ce{Pd} served as the top electrode~\cite{JoFr2020}.
Fabrication process is described in more detail in~\cite{LiBe2018}.

\threesubsection{Characterization}

Device conductance was being increased using SET pulses (\SI{500}{\micro\second} @ \SI{2.5}{\volt} and gate voltage linearly increasing from \SI{0.6}{\volt} to \SI{1.6}{\volt}).
After each \num{100}-pulse cycle, RESET pulses (\SI{5}{\micro\second} @ \SI{0.9}{\volt} linearly increasing to \SI{2.2}{\volt} and gate voltage of \SI{5}{\volt}) were used to reduce the conductance.
More information can be found in~\cite{WaLi2019}.

\subsection{Simulations}

\subsubsection{Training and inference particulars}

The following architectures were employed:
\begin{itemize}
    \item fully connected \glspl{ANN} (trained on MNIST~\cite{LeCo2010}) containing
    \begin{enumerate}
        \item fully connected layer with \num{25} hidden neurons and logistic activation function
        \item fully connected layer with \num{10} output neurons and softmax activation function
    \end{enumerate}
    \item \glspl{CNN} (trained on CIFAR-10~\cite{Kr2009}) containing
    \begin{enumerate}
        \item convolutional layer with \num{32} output filters, $3 \times 3$ kernel size and ReLU activation function
        \item pooling layer with $2 \times 2$ pool size
        \item convolutional layer with \num{64} output filters, $3 \times 3$ kernel size and ReLU activation function
        \item pooling layer with $2 \times 2$ pool size
        \item convolutional layer with \num{64} output filters, $3 \times 3$ kernel size and ReLU activation function with maximum value of \num{1}
        \item fully connected layer with \num{25} hidden neurons and logistic activation function
        \item fully connected layer with \num{10} output neurons and softmax activation function
    \end{enumerate}
\end{itemize}

To account for high variability of nonidealities (which were nondeterministic), \num{5} networks were trained for each configuration.
Each trained network went through \num{25} inference runs, totaling $5 \times 25 = 125$ runs for each configuration.

Networks used $4:1$ training-validation split.
All networks were trained for a \num{1000} epochs with batch size of \num{64}.
Where $\ell_1$ regularization had been employed, regularization factor of $10^{-4}$ was used.
To ensure double weights stayed nonnegative we utilized \inlinecode{NonNeg} weight constraint provided by the Keras machine learning library. 
For any given batch (whose size was \num{64} during training, as mentioned before, and \num{100} during inference), conductances were disturbed \emph{once} in the case of linearity-preserving nonidealities and Poole-Frenkel parameters associated with individual devices were drawn from a probability distribution \emph{once} in the case of \IV\ nonlinearity.

\subsection{Statistical Analysis}\label{subsec:statistical-analysis}

\subsubsection{Data presentation}

\begin{itemize}
  \item In all box plots, the maximum whisker length is set to $1.5 \times \mathrm{IQR}$.
  \item In \cref{fig:iv-nonlinearity-training} and equivalent plots, curves with semitransparent regions consist of two parts summarizing \num{20} inference repeats at certain epochs: opaque curve representing the median values and semi-transparent region bounded by the minimum and maximum values.
  \item To avoid large file size, only a subset of all data points is presented in \cref{fig:Ta-HfO2,fig:weight-implementation}; these subsets were chosen randomly using NumPy.
\end{itemize}

\subsubsection{Experimental \ce{SiO_x} data}

\num{53} \ce{SiO_x} resistance states were achieved using the procedure described in \cref{sec:SiO_x-experiments} but several of them were excluded from the analysis.
Specifically, states where there were abrupt changes in current were not considered.
This was done by excluding the curves where the maximum ratio of the second derivative of current (with respect to voltage) to average current exceeded a threshold of \SI{0.1}{\volt^{-2}}.

\begin{figure*}[t!]
  \begin{center}
    \includegraphics{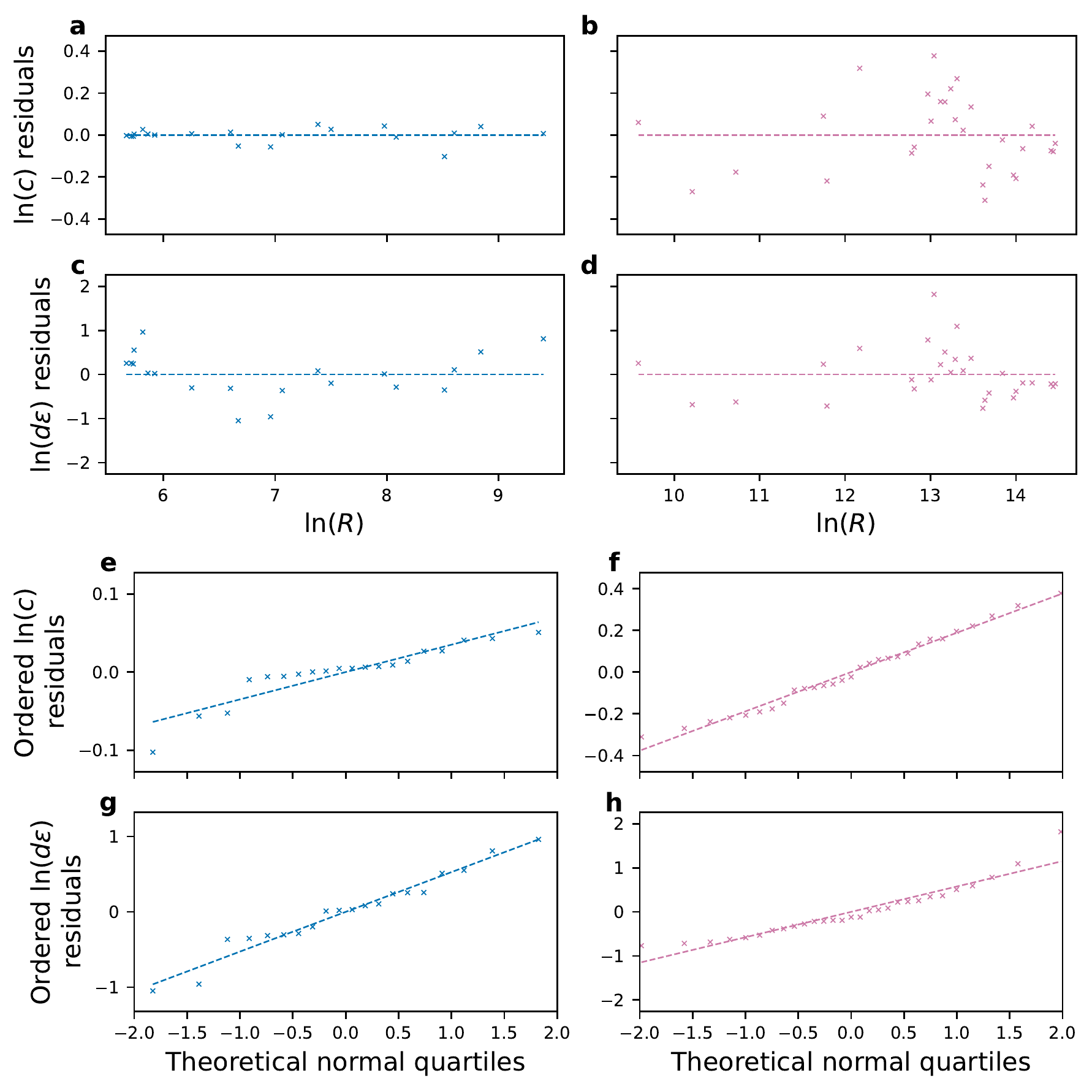}
    {\phantomsubcaption\label{fig:pf-residuals:a}}
    {\phantomsubcaption\label{fig:pf-residuals:b}}
    {\phantomsubcaption\label{fig:pf-residuals:c}}
    {\phantomsubcaption\label{fig:pf-residuals:d}}
    {\phantomsubcaption\label{fig:pf-residuals:e}}
    {\phantomsubcaption\label{fig:pf-residuals:f}}
    {\phantomsubcaption\label{fig:pf-residuals:g}}
    {\phantomsubcaption\label{fig:pf-residuals:h}}
  \end{center}
  \mycaption{Residuals from the trends of Poole-Frenkel parameters}{%
    Residuals of $\ln(c)$ for \subf{a}) low- and \subf{b}) high-resistance states.
    Residuals of $\ln(d \varepsilon)$ for \subf{c}) low- and \subf{d}) high-resistance states.
    Normal probability plots of the residuals of $\ln(c)$ for \subf{e}) low- and \subf{f}) high-resistance states.
    Normal probability plots of the residuals of $\ln(d \varepsilon)$ for \subf{g}) low- and \subf{h}) high-resistance states.
    In all panels, the inputs to logarithms are made dimensionless by using the amounts of the corresponding quantities in SI units.
  }\label{fig:pf-residuals}
\end{figure*}

The analysis of the residuals of $\ln(c)$ and $\ln(d \varepsilon)$ is provided in \cref{fig:pf-residuals}.
One of the issues, which was evident, was that these two sets of residuals correlated to some extent, especially at higher resistance states as can be seen in \cref{fig:pf-residuals:b,fig:pf-residuals:b}.
If these deviations were simulated independently, the amount of uncertainty would be significantly overestimated, which is why covariance matrix of the residuals is used in \cref{eq:g-iv-nonlinearity}.
The rationale for simulating the deviations using normal distribution is provided in the normal probability plots in \cref{fig:pf-residuals:e,fig:pf-residuals:f,fig:pf-residuals:g,fig:pf-residuals:h}.

\vspace{3in}

\section*{Acknowledgments}

A.\,M.\ acknowledges funding from the Royal Academy of Engineering under the Research Fellowship scheme, G.\,A.\,C.\ acknowledges funding from the Engineering and Physical Sciences Research Council (EP/S030069/1 and EP/L016796/1), A.\,J.\,K.\ acknowledges funding from the Engineering and Physical Sciences Research Council (EP/P013503/1) and the Leverhulme Trust (RPG-2016-135), D.\,J.\ acknowledges studentship funding from the Engineering and Physical Sciences Research Council (ref.\ 2094654).

Authors thank Professor Qiangfei Xia of University of Massachusetts Amherst for sharing data on the switching of \ce{Ta}/\ce{HfO2} devices.

\section*{Conflict of Interest}

A.\,M.\ and A.\,J.\,K.\ are co-founders and W.\,H.\,N.\ is an employee of Intrinsic, a company developing memristor technology.

\section*{Author Contributions}

D.\,J.\ and E.\,W.\ contributed equally to this work, as did A.\,M.\ and G.\,A.\,C.

\section*{Data Availability Statement}

The source code for all the simulations in the paper can be found at \url{https://github.com/joksas/nonideality-aware-mnn-training/releases/tag/2022-04-22}.
\ce{Ta}/\ce{HfO2} data that support the findings of this study are available from D.\,J.\ upon request.
\ce{SiO_x} data have been made publicly available~\cite{BaMe2021}; file \texttt{excelDataCombined.mat} was utilized in the simulations of this work.
All data generated by the simulations have been made publicly available~\cite{JoWa2021}.

\clearpage
\printbibliography

\clearpage

\newgeometry{onecolumn, hmargin=15mm, bottom=20mm, top=25mm}

\makeatletter
\renewcommand\thefigure{S\arabic{figure}}
\renewcommand\thetable{S\arabic{table}}
\makeatother
\setcounter{table}{0}
\setcounter{figure}{0}

\section*{Supporting Tables}

\begin{table}[h!]
  \centering
  \begin{tabular}{clccl}
    ID & Nonideality & $G_\mathrm{off}$ (\si{\siemens}) & $G_\mathrm{on}$ (\si{\siemens}) & Parameters \\
    \midrule
    \labelni{ni:low-iv-nonlinearity} & Low \IV\ nonlinearity in \ce{SiO_x} & \num{6.901e-4} & \num{3.451e-3} & See \crefExt{fig:SiO_x} \\
    \labelni{ni:high-iv-nonlinearity} & High \IV\ nonlinearity in \ce{SiO_x} & \num{5.248e-7} & \num{2.624e-6} & See \crefExt{fig:SiO_x} \\
    \labelni{ni:stuck-off} & Stuck at $G_\mathrm{off}$ & - & - & $P(\text{stuck}) = 0.05$ \\
    \labelni{ni:stuck-on} & Stuck at $G_\mathrm{on}$ & - & - & $P(\text{stuck}) = 0.05$ \\
    \labelni{ni:stuck-distribution} & Stuck \ce{Ta}/\ce{HfO2} devices & \num{4.364e-5} & \num{9.782e-4} & $P(\text{stuck}) = 0.101$ \\
    \labelni{ni:more-uniform-d2d} & More uniform \glsentryshort{D2D} variability & - & - & $\sigma|_{R=R_\mathrm{off}} = \sigma|_{R=R_\mathrm{on}} = 0.25$ \\
    \labelni{ni:less-uniform-d2d} & Less uniform \glsentryshort{D2D} variability & - & - & $\sigma|_{R=R_\mathrm{off}} = 0.5$, $\sigma|_{R=R_\mathrm{on}} = 0.05$ \\
    \labelni{ni:high-d2d} & High-magnitude \glsentryshort{D2D} variability & - & - & $\sigma|_{R=R_\mathrm{off}} = \sigma|_{R=R_\mathrm{on}} = 0.5$ \\
  \end{tabular}
  \mycaption{All nonidealities utilized in the simulations}{For mappings onto crossbar arrays, $G_\mathrm{off}$ and $G_\mathrm{on}$ of the nonidealities were used. Where these values were not available (because the models did not use experimental data), they were borrowed from \cref{ni:high-iv-nonlinearity}.}\label{supp-tab:nonidealities}
\end{table}

\begin{table}[h!]
  \centering
  \begin{tabular}{clll}
    ID & Nonideality IDs & Notes & Figures \\
    \midrule
    \labeltr{tr:ideal} & - & Mapping from \crefExt{eq:mapping-low-G} and standard validation. & \labelcrefExt{fig:iv-nonlinearity-training:a,fig:iv-nonlinearity-training:d,fig:iv-nonlinearity-inference,fig:iv-nonlinearity-cnn:c,fig:nonideality-agnosticism} \\
    \labeltr{tr:low-iv-nonlinearity} & \labelcref{ni:low-iv-nonlinearity} & - & \labelcrefExt{fig:iv-nonlinearity-training:b,fig:iv-nonlinearity-inference,fig:nonideality-agnosticism} \\
    \labeltr{tr:low-iv-nonlinearity-reg} & \labelcref{ni:low-iv-nonlinearity} & Regularized. & \labelcrefExt{fig:iv-nonlinearity-training:c,fig:iv-nonlinearity-inference,fig:nonideality-agnosticism} \\
    \labeltr{tr:high-iv-nonlinearity} & \labelcref{ni:high-iv-nonlinearity} & - & \labelcrefExt{fig:iv-nonlinearity-training:e,fig:iv-nonlinearity-inference,fig:iv-nonlinearity-cnn,fig:nonideality-agnosticism} \\
    \labeltr{tr:high-iv-nonlinearity-reg} & \labelcref{ni:high-iv-nonlinearity} & Regularized. & \labelcrefExt{fig:iv-nonlinearity-training:f,fig:iv-nonlinearity-inference,fig:nonideality-agnosticism} \\
    \labeltr{tr:more-uniform-d2d-symmetric-G} & \labelcref{ni:more-uniform-d2d} & Mapping from \crefExt{eq:mapping-symmetric-G}. & \labelcrefExt{fig:weight-implementation:a,fig:weight-implementation:i} \\
    \labeltr{tr:more-uniform-d2d-low-G} & \labelcref{ni:more-uniform-d2d} & Mapping from \crefExt{eq:mapping-low-G}. & \labelcrefExt{fig:weight-implementation:b,fig:weight-implementation:i} \\
    \labeltr{tr:more-uniform-d2d} & \labelcref{ni:more-uniform-d2d} & - & \labelcrefExt{fig:weight-implementation:c,fig:weight-implementation:i,fig:nonideality-agnosticism} \\
    \labeltr{tr:more-uniform-d2d-reg} & \labelcref{ni:more-uniform-d2d} & Regularized. & \labelcrefExt{fig:weight-implementation:d,fig:weight-implementation:i,fig:nonideality-agnosticism} \\
    \labeltr{tr:less-uniform-d2d-symmetric-G} & \labelcref{ni:less-uniform-d2d} & Mapping from \crefExt{eq:mapping-symmetric-G}. & \labelcrefExt{fig:weight-implementation:e,fig:weight-implementation:j} \\
    \labeltr{tr:less-uniform-d2d-low-G} & \labelcref{ni:less-uniform-d2d} & Mapping from \crefExt{eq:mapping-low-G}. & \labelcrefExt{fig:weight-implementation:f,fig:weight-implementation:j} \\
    \labeltr{tr:less-uniform-d2d} & \labelcref{ni:less-uniform-d2d} & - & \labelcrefExt{fig:weight-implementation:g,fig:weight-implementation:j,fig:nonideality-agnosticism} \\
    \labeltr{tr:less-uniform-d2d-reg} & \labelcref{ni:less-uniform-d2d} & Regularized. & \labelcrefExt{fig:weight-implementation:h,fig:weight-implementation:j,fig:nonideality-agnosticism} \\
    \labeltr{tr:high-d2d} & \labelcref{ni:high-d2d} & - & \labelcrefExt{fig:nonideality-agnosticism} \\
    \labeltr{tr:stuck-off} & \labelcref{ni:stuck-off} & - & \labelcrefExt{fig:nonideality-agnosticism} \\
    \labeltr{tr:stuck-distribution} & \labelcref{ni:stuck-distribution} & - & \labelcrefExt{fig:nonideality-agnosticism} \\
    \labeltr{tr:high-iv-nonlinearity-and-stuck-on} & \labelcref{ni:high-iv-nonlinearity,ni:stuck-on} & - & \labelcrefExt{fig:nonideality-agnosticism} \\
  \end{tabular}
  \mycaption{Training setups}{Unless stated otherwise, the networks used double weights, regularization was not applied and memristive validation was used.}\label{supp-tab:training-setups}
\end{table}

\begin{table}[h!]
  \centering
  \begin{tabular}{clll}
    ID & Nonideality IDs & Notes & Figures \\
    \midrule
    \labelinf{inf:ideal} & - & Mapping from \crefExt{eq:mapping-low-G}. & \labelcrefExt{fig:nonideality-agnosticism} \\
    \labelinf{inf:low-iv-nonlinearity} & \labelcref{ni:low-iv-nonlinearity} & - & \labelcrefExt{fig:iv-nonlinearity-training:a,fig:iv-nonlinearity-training:b,fig:iv-nonlinearity-training:c,fig:iv-nonlinearity-inference,fig:nonideality-agnosticism} \\
    \labelinf{inf:high-iv-nonlinearity} & \labelcref{ni:high-iv-nonlinearity} & - & \labelcrefExt{fig:iv-nonlinearity-training:d,fig:iv-nonlinearity-training:e,fig:iv-nonlinearity-training:f,fig:iv-nonlinearity-inference,fig:iv-nonlinearity-cnn,fig:nonideality-agnosticism} \\
    \labelinf{inf:more-uniform-d2d-symmetric-G} & \labelcref{ni:more-uniform-d2d} & Mapping from \crefExt{eq:mapping-symmetric-G}. & \labelcrefExt{fig:weight-implementation:i} \\
    \labelinf{inf:more-uniform-d2d-low-G} & \labelcref{ni:more-uniform-d2d} & Mapping from \crefExt{eq:mapping-low-G}. & \labelcrefExt{fig:weight-implementation:i} \\
    \labelinf{inf:more-uniform-d2d} & \labelcref{ni:more-uniform-d2d} & - & \labelcrefExt{fig:weight-implementation:i,fig:nonideality-agnosticism} \\
    \labelinf{inf:less-uniform-d2d-symmetric-G} & \labelcref{ni:less-uniform-d2d} & Mapping from \crefExt{eq:mapping-symmetric-G}. & \labelcrefExt{fig:weight-implementation:j} \\
    \labelinf{inf:less-uniform-d2d-low-G} & \labelcref{ni:less-uniform-d2d} & Mapping from \crefExt{eq:mapping-low-G}. & \labelcrefExt{fig:weight-implementation:j} \\
    \labelinf{inf:less-uniform-d2d} & \labelcref{ni:less-uniform-d2d} & - & \labelcrefExt{fig:weight-implementation:j,fig:nonideality-agnosticism} \\
    \labelinf{inf:high-d2d} & \labelcref{ni:high-d2d} & - & \labelcrefExt{fig:nonideality-agnosticism} \\
    \labelinf{inf:stuck-off} & \labelcref{ni:stuck-off} & - & \labelcrefExt{fig:nonideality-agnosticism} \\
    \labelinf{inf:stuck-distribution} & \labelcref{ni:stuck-distribution} & - & \labelcrefExt{fig:nonideality-agnosticism} \\
    \labelinf{inf:high-iv-nonlinearity-and-stuck-on} & \labelcref{ni:high-iv-nonlinearity,ni:stuck-on} & - & \labelcrefExt{fig:nonideality-agnosticism} \\
  \end{tabular}
  \mycaption{Inference/test setups}{This includes both the figures where the networks were evaluated on the test set after they had been fully trained and the figures in which test set performance was evaluated during training. Unless stated otherwise, the networks used double weights.}\label{supp-tab:inference-setups}
\end{table}

\begin{table}[H]
  \centering
  \begin{tabular}{lcc}
    Name & Training setup ID & Inference setup ID \\
    \midrule
    Ideal & \labelcref{tr:ideal} & \labelcref{inf:ideal} \\
    Low $I$-$V$ nonlin. [$\mathrm{SiO}_x$] & \labelcref{tr:low-iv-nonlinearity} & \labelcref{inf:low-iv-nonlinearity} \\
    Low $I$-$V$ nonlin. [$\mathrm{SiO}_x$] (reg.) & \labelcref{tr:low-iv-nonlinearity-reg} & - \\
    High $I$-$V$ nonlin. [$\mathrm{SiO}_x$] & \labelcref{tr:high-iv-nonlinearity} & \labelcref{inf:high-iv-nonlinearity} \\
    High $I$-$V$ nonlin. [$\mathrm{SiO}_x$] (reg.) & \labelcref{tr:high-iv-nonlinearity-reg} & - \\
    Stuck at $G_\mathrm{off}$ & \labelcref{tr:stuck-off} & \labelcref{inf:stuck-off} \\
    Stuck [$\mathrm{Ta/HfO}_2$] & \labelcref{tr:stuck-distribution} & \labelcref{inf:stuck-distribution} \\
    More uniform D2D var. & \labelcref{tr:more-uniform-d2d} & \labelcref{inf:more-uniform-d2d} \\
    More uniform D2D var. (reg.) & \labelcref{tr:more-uniform-d2d-reg} & - \\
    Less uniform D2D var. & \labelcref{tr:less-uniform-d2d} & \labelcref{inf:less-uniform-d2d} \\
    Less uniform D2D var. (reg.) & \labelcref{tr:less-uniform-d2d-reg} & - \\
    High $I$-$V$ nonlin. [$\mathrm{SiO}_x$] + stuck at $G_\mathrm{on}$ & \labelcref{tr:high-iv-nonlinearity-and-stuck-on} & \labelcref{inf:high-iv-nonlinearity-and-stuck-on} \\
    High D2D var. & \labelcref{tr:high-d2d} & \labelcref{inf:high-d2d}
  \end{tabular}
  \mycaption{Training and inference setups used in \crefExt{fig:nonideality-agnosticism}}{}\label{supp-tab:figure-7-setups}
\end{table}

\clearpage

\section*{Supporting Figures}

\begin{figure}[H]
  \begin{center}
    \includegraphics{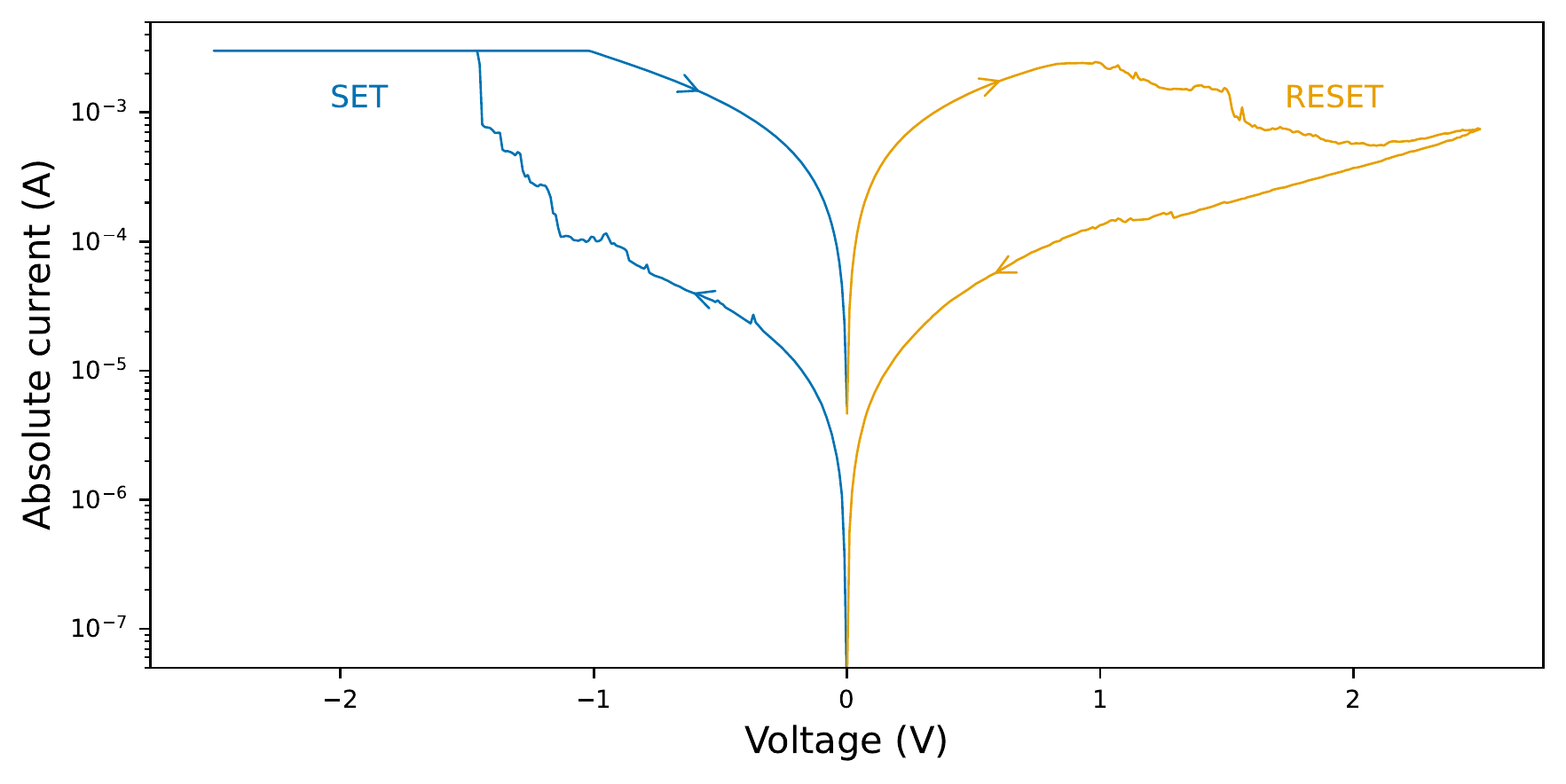}
  \end{center}
  \mycaption{Typical resistance switching behavior of the \ce{SiO_x} device}{%
    One SET and one RESET sweep are shown.
  }\label{supp-fig:SiO_x-switching}
\end{figure}

\begin{figure}[H]
  \begin{center}
    \includegraphics{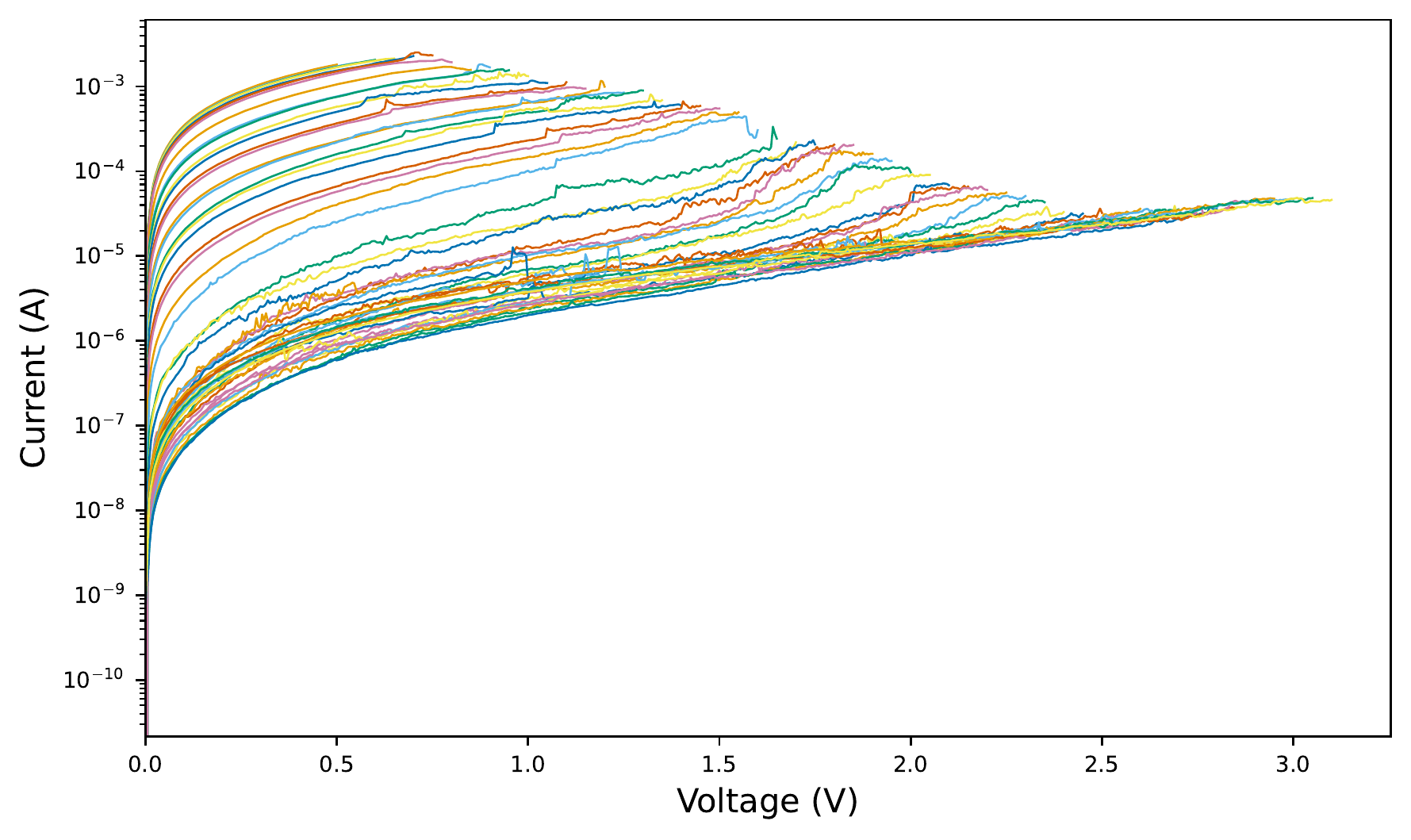}
  \end{center}
  \mycaption{All \IV\ sweeps of the \ce{SiO_x} device}{%
    Single sweeps are shown for \num{53} different states obtained by incrementing maximum voltage by \SI{0.05}{\volt}.
    Every seventh state shares the same color.
  }\label{supp-fig:all-SiO_x-IV-curves-full-range}
\end{figure}

\begin{figure}[H]
  \begin{center}
    \includegraphics{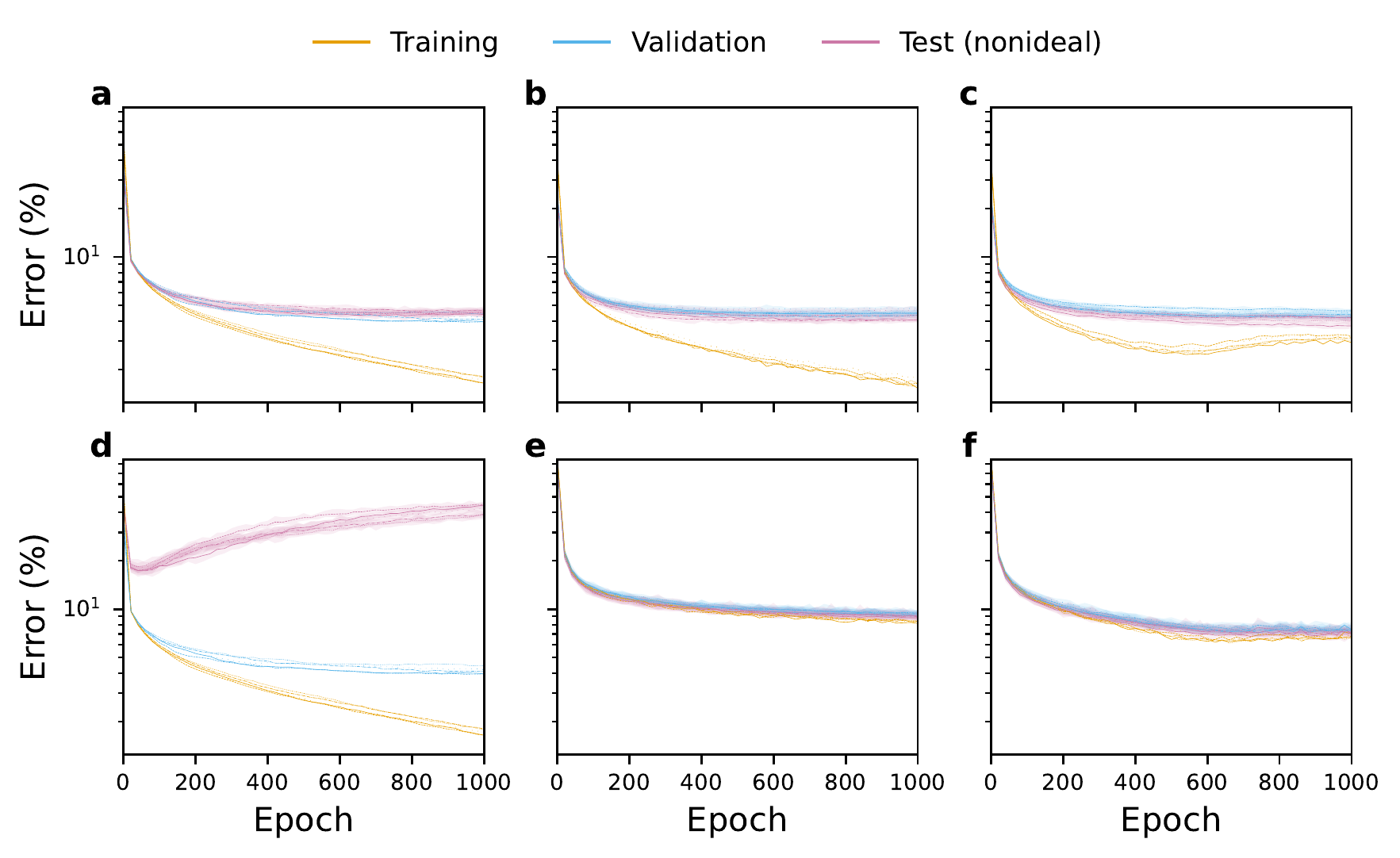}
  \end{center}
  \mycaption{Training results for standard and nonideality-aware schemes when exposed to \IV\ nonlinearities}{%
    Panels include equivalent curves of \crefExt{fig:iv-nonlinearity-training} for all five sets of trained networks.
    \subf{a},\,\subf{d}) \Cref{tr:ideal}, \subf{b}) \cref{tr:low-iv-nonlinearity}, \subf{c}) \cref{tr:low-iv-nonlinearity-reg}, \subf{e}) \cref{tr:high-iv-nonlinearity}, \subf{f}) \cref{tr:high-iv-nonlinearity-reg}; \subf{a}--\subf{c}) \cref{inf:low-iv-nonlinearity}, \subf{d}--\subf{f}) \cref{inf:high-iv-nonlinearity}.
    Curves of different networks are indicated by different line styles in each of the panels.
  }\label{supp-fig:iv-nonlinearity-training}
\end{figure}

\begin{figure}[H]
  \begin{center}
    \includegraphics{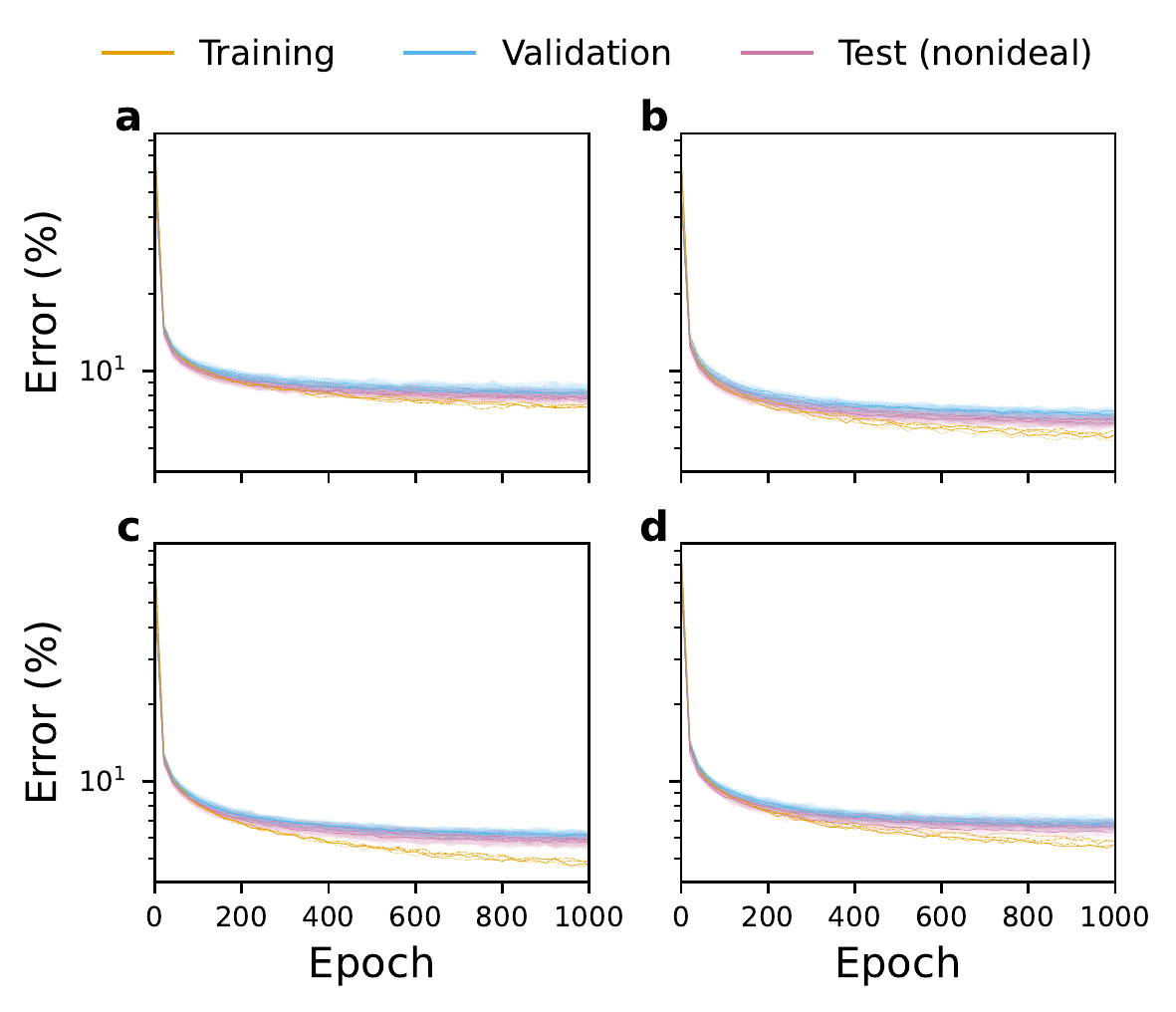}
  \end{center}
  \mycaption{Training results for nonideality-aware scheme with conventional weight implementations when exposed to \glsentrylong{D2D} variability}{%
    \subf{a}) \Cref{tr:more-uniform-d2d-symmetric-G}, \cref{inf:more-uniform-d2d-symmetric-G}, \subf{b}) \cref{tr:more-uniform-d2d-low-G}, \cref{inf:more-uniform-d2d-low-G}, \subf{c}) \cref{tr:less-uniform-d2d-symmetric-G}, \cref{inf:less-uniform-d2d-symmetric-G}, \subf{d}) \cref{tr:less-uniform-d2d-low-G}, \cref{inf:less-uniform-d2d-low-G}.
    Curves of different networks are indicated by different line styles in each of the panels.
  }\label{supp-fig:weight-implementation-standard-weights-training}
\end{figure}

\begin{figure}[H]
  \begin{center}
    \includegraphics{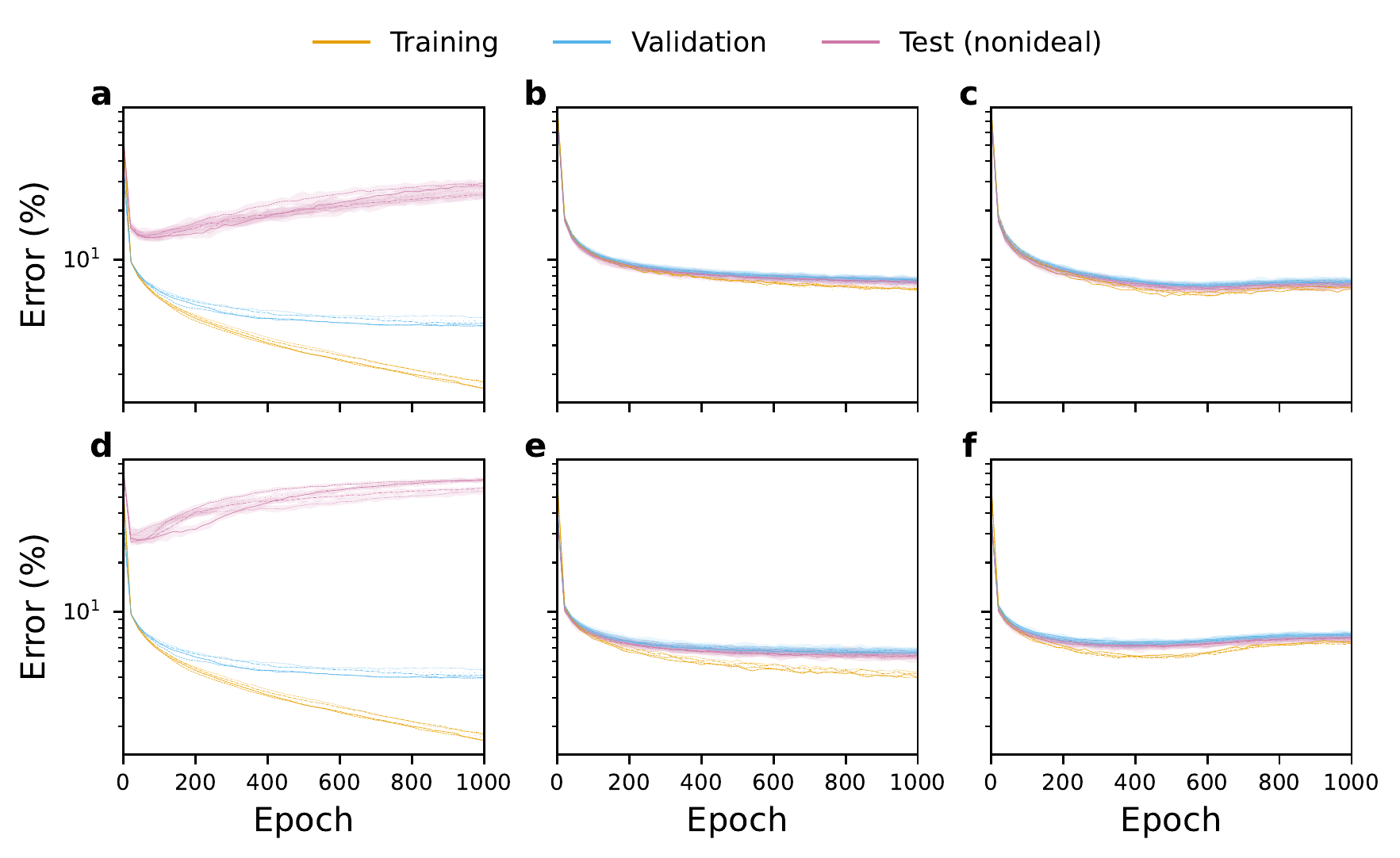}
  \end{center}
  \mycaption{Training results for standard and nonideality-aware schemes when exposed to \glsentrylong{D2D} variability}{%
    \subf{a},\,\subf{d}) \Cref{tr:ideal}, \subf{b}) \cref{tr:more-uniform-d2d}, \subf{c}) \cref{tr:more-uniform-d2d-reg}, \subf{e}) \cref{tr:less-uniform-d2d}, \subf{f}) \cref{tr:less-uniform-d2d-reg}; \subf{a}--\subf{c}) \cref{inf:more-uniform-d2d}, \subf{d}--\subf{f}) \cref{inf:less-uniform-d2d}.
    Curves of different networks are indicated by different line styles in each of the panels.
  }\label{supp-fig:weight-implementation-double-weights-training}
\end{figure}

\begin{figure}[H]
  \begin{center}
    \includegraphics{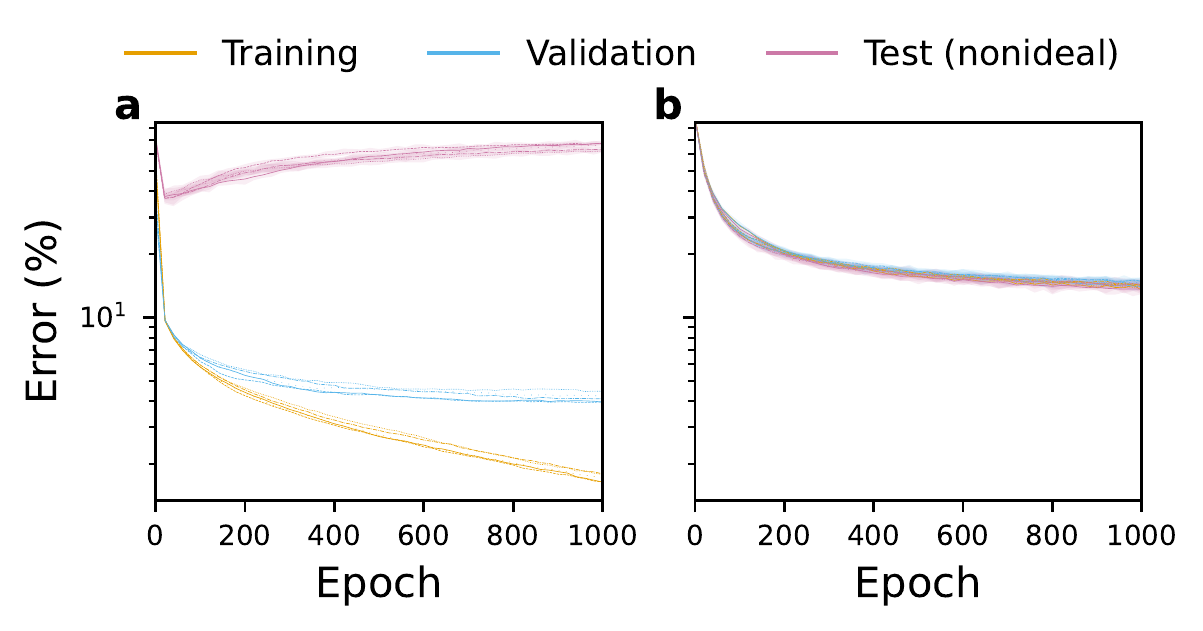}
  \end{center}
  \mycaption{Training results for standard and nonideality-aware schemes when exposed to high-magnitude \glsentrylong{D2D} variability}{%
    \subf{a}) \Cref{tr:ideal}, \subf{b}) \cref{tr:high-d2d}; \cref{inf:high-d2d} was used in both panels.
    Curves of different networks are indicated by different line styles in each of the panels.
    Highly stochastic nonidealities like high-magnitude \glsentrylong{D2D} variability prevent the training from overfitting to a particular set of behaviors---this is evident from how coupled training, validation and test curves are in \subf{b}.
  }\label{supp-fig:high-d2d-training}
\end{figure}

\begin{figure}[H]
  \begin{center}
    \includegraphics{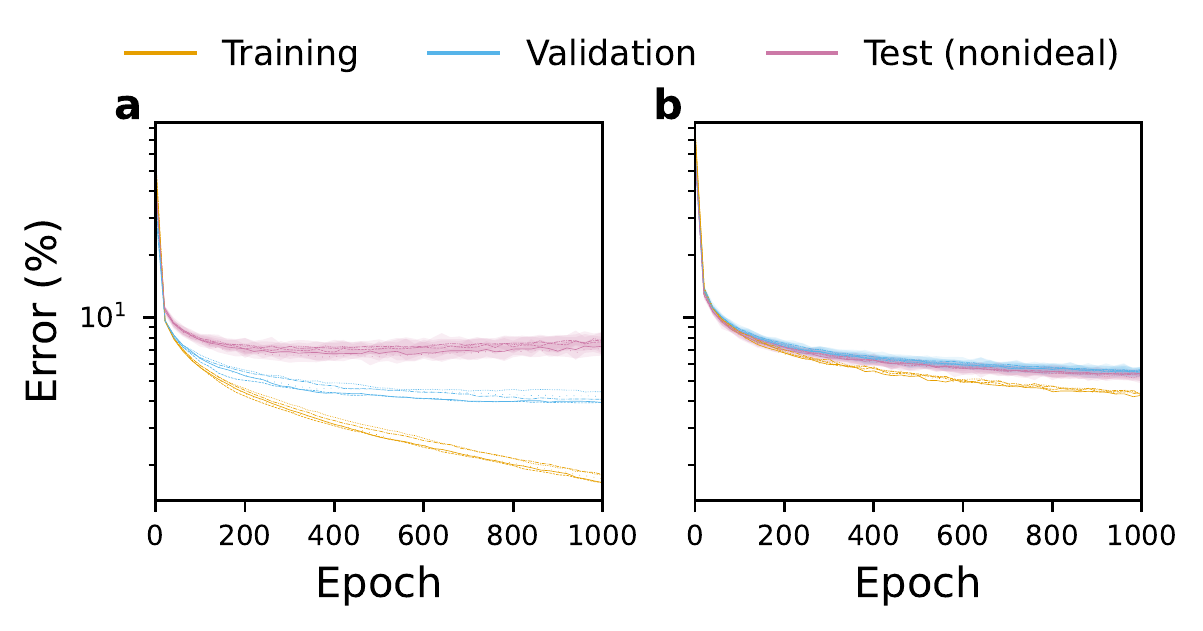}\
  \end{center}
  \mycaption{Training results for standard and nonideality-aware schemes when exposed to devices getting stuck at $G_\mathrm{off}$}{%
    \subf{a}) \Cref{tr:ideal}, \subf{b}) \cref{tr:stuck-off}; \cref{inf:stuck-off} was used in both panels.
    Curves of different networks are indicated by different line styles in each of the panels.
  }\label{supp-fig:stuck-off-training}
\end{figure}

\begin{figure}[H]
  \begin{center}
    \includegraphics{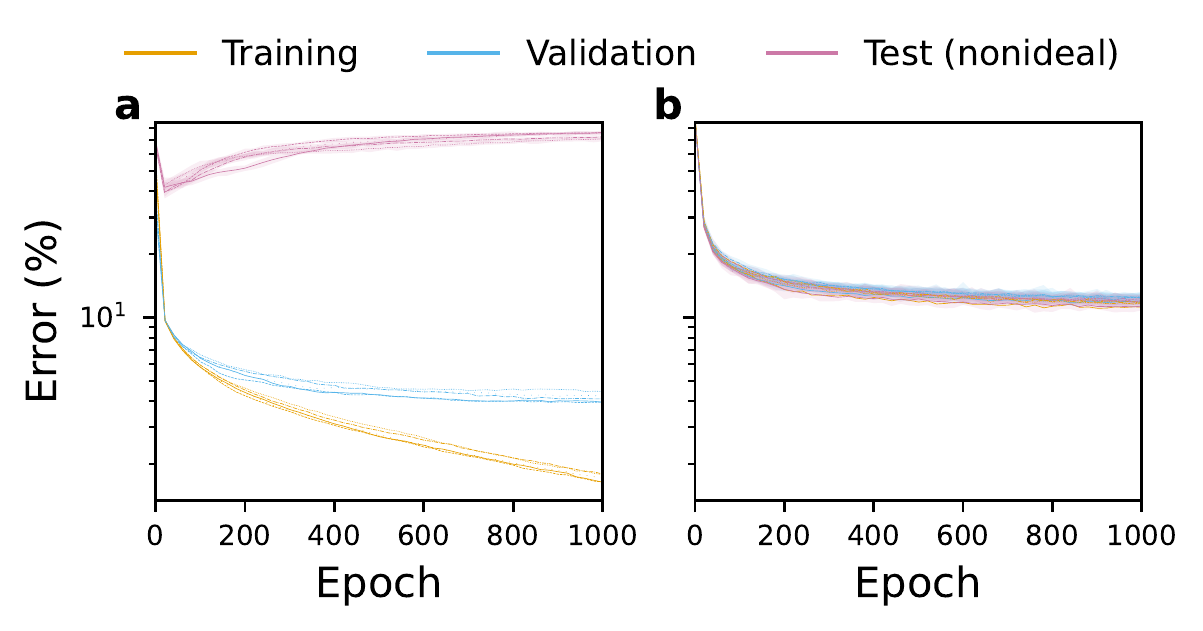}\
  \end{center}
  \mycaption{Training results for standard and nonideality-aware schemes when exposed to high \IV\ nonlinearity and devices getting stuck at $G_\mathrm{on}$}{%
    \subf{a}) \Cref{tr:ideal}, \subf{b}) \cref{tr:high-iv-nonlinearity-and-stuck-on}; \cref{inf:high-iv-nonlinearity-and-stuck-on} was used in both panels.
    Curves of different networks are indicated by different line styles in each of the panels.
  }\label{supp-fig:high-iv-nonlinearity-and-stuck-on-training}
\end{figure}

\begin{figure}[H]
  \begin{center}
    \includegraphics{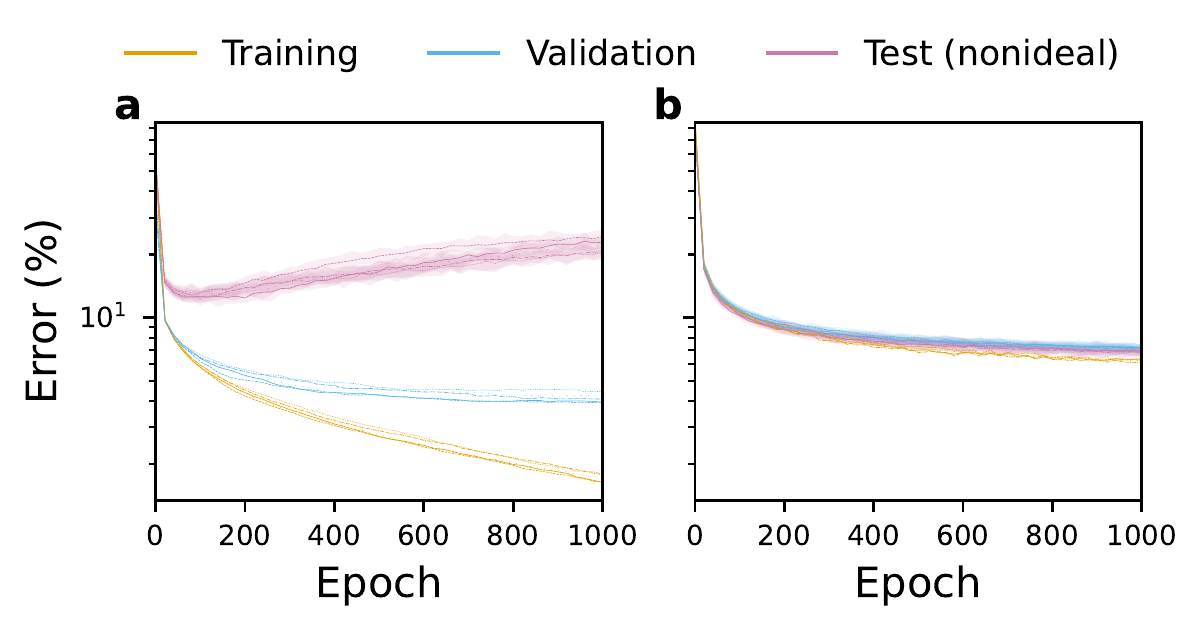}\
  \end{center}
  \mycaption{Training results for standard and nonideality-aware schemes when exposed to devices getting stuck}{%
    \subf{a}) \Cref{tr:ideal}, \subf{b}) \cref{tr:stuck-distribution}; \cref{inf:stuck-distribution} was used in both panels.
    Curves of different networks are indicated by different line styles in each of the panels.
  }\label{supp-fig:stuck-distribution-training}
\end{figure}

\begin{figure}[H]
  \begin{center}
    \includegraphics{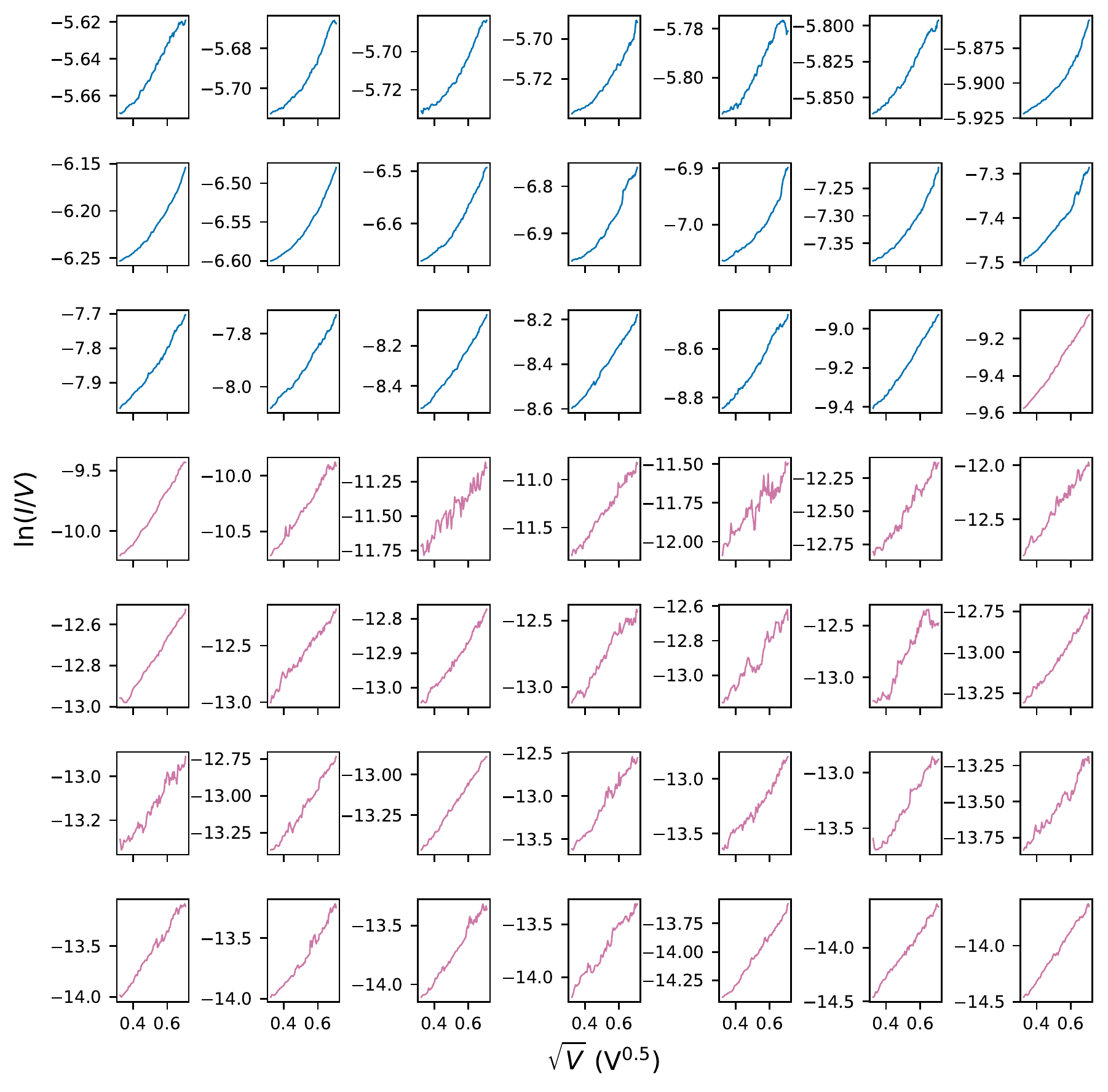}
  \end{center}
  \mycaption{Poole-Frenkel plots for \num{49} resistance states}{%
    The plots are shown for both low-resistance (in blue) and high-resistance (in pink) states in the range from \SI{0.1}{\volt} to \SI{0.5}{\volt}.
    The inputs to logarithms are made dimensionless by using the amounts of the corresponding quantities in SI units.
  }\label{supp-fig:pf-plots}
\end{figure}

\begin{figure}[H]
  \begin{center}
    \includegraphics{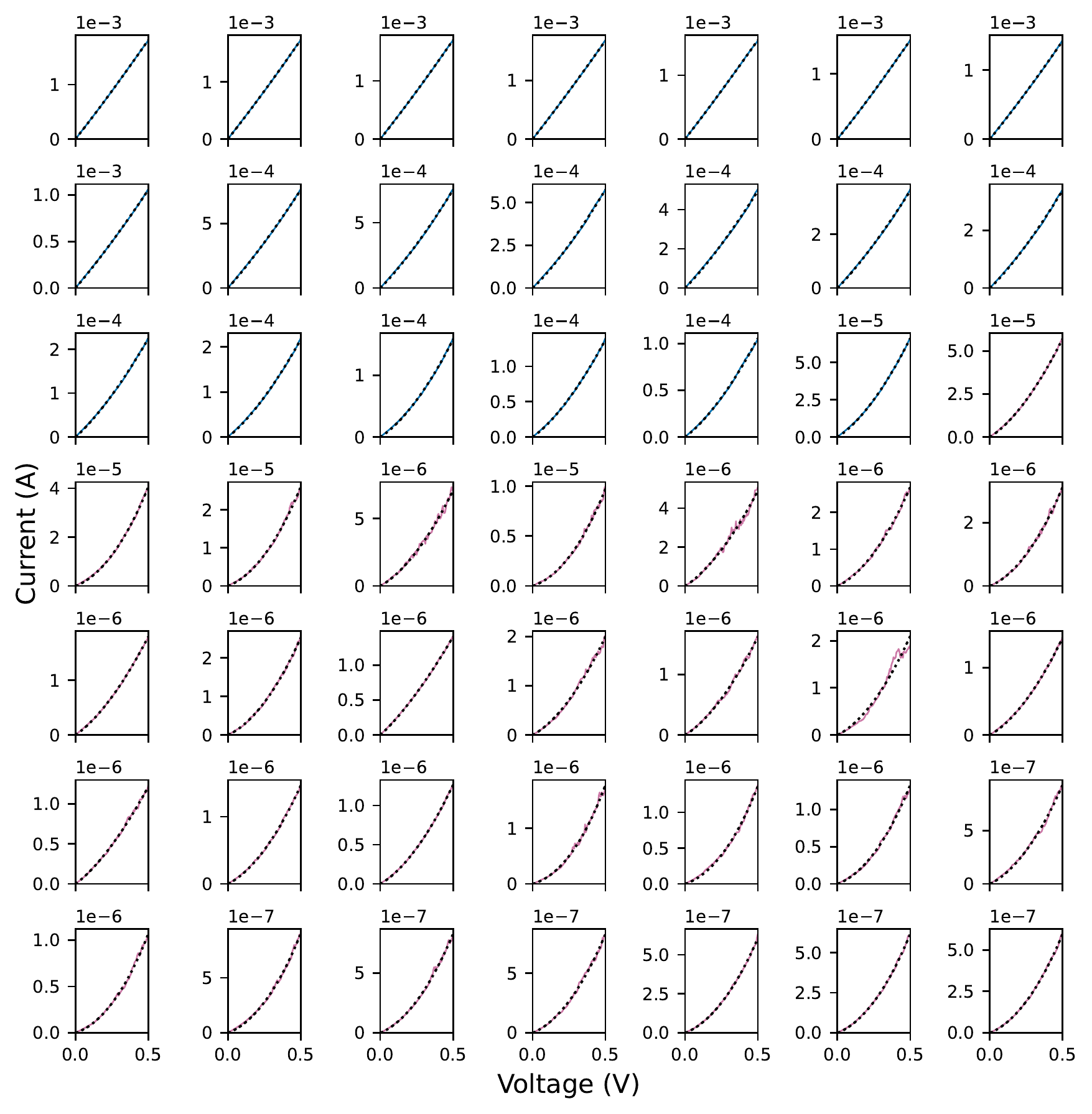}
  \end{center}
  \mycaption{Poole-Frenkel fits for \num{49} resistance states}{%
    The fits (as black dotted curves) are drawn over the \IV\ curves of both low-resistance (in blue) and high-resistance (in pink) states.
  }\label{supp-fig:pf-fits}
\end{figure}

\end{document}